\documentclass[twocolumn]{aastex631}

\usepackage{xspace,upgreek}

\newcommand{\rtwelve}{FRB~20190208A\xspace}
\newcommand{\rnarwhal}{FRB~20220912A\xspace}
\newcommand{\rone}{FRB~20121102A\xspace}
\newcommand{\rthree}{FRB~20180916B\xspace}

\newcommand{\rsixtyseven}{FRB~20201124A\xspace}
\newcommand{\ronetwin}{FRB~20190520B\xspace}

\begin{document}

\title{A repeating fast radio burst source in a low-luminosity dwarf galaxy}

\author[0000-0002-0786-7307]{Dant\'e~M. Hewitt}
\affiliation{Anton Pannekoek Institute for Astronomy, University of Amsterdam, Science Park 904, 1098~XH Amsterdam, the Netherlands}

\author[0000-0002-3615-3514]{Mohit Bhardwaj}
\affiliation{McWilliams Center for Cosmology, Department of Physics, Carnegie Mellon University, Pittsburgh, PA 15213, USA}

\author[0000-0002-5025-4645]{Alexa~C. Gordon}
\affiliation{Center for Interdisciplinary Exploration and Research in Astrophysics (CIERA) and Department of Physics and Astronomy, Northwestern University, Evanston, IL 60208, USA}

\author[0000-0002-8139-8414]{Aida Kirichenko}
\affiliation{Instituto de Astronomía, Universidad Nacional Autónoma de México, Apdo. Postal 877, Ensenada, Baja California 22800, México}

\author[0000-0003-0510-0740]{Kenzie Nimmo}
\affiliation{MIT Kavli Institute for Astrophysics and Space Research, Massachusetts Institute of Technology, 77 Massachusetts Ave, Cambridge, MA 02139, USA}

\author[0000-0003-3460-506X]{Shivani Bhandari}
\affiliation{ASTRON, Netherlands Institute for Radio Astronomy, Oude Hoogeveensedijk 4, 7991~PD Dwingeloo, the Netherlands}
\affiliation{Joint Institute for VLBI ERIC, Oude Hoogeveensedijk 4, 7991~PD Dwingeloo, the Netherlands}
\affiliation{Anton Pannekoek Institute for Astronomy, University of Amsterdam, Science Park 904, 1098~XH Amsterdam, the Netherlands}
\affiliation{CSIRO Space and Astronomy, Australia Telescope National Facility, PO Box 76, Epping, NSW~1710, Australia}

\author[0000-0002-1775-9692]{Isma\"el Cognard}
\affiliation{Observatoire Radioastronomique de Nan\c{c}ay, Observatoire de Paris, Universit\'e PSL, Universit\'e d’Orléans, CNRS, 18330 Nan\c{c}ay, France}

\author[0000-0002-7374-935X]{Wen-fai Fong}
\affiliation{Center for Interdisciplinary Exploration and Research in Astrophysics (CIERA) and Department of Physics and Astronomy, Northwestern University, Evanston, IL 60208, USA}

\author[0000-0001-6150-2854]{Armando Gil de Paz}
\affiliation{Departamento de Física de la Tierra y Astrofísica, Facultad de CC. Físicas, Universidad Complutense de Madrid, E-28040, Madrid, Spain}
\affiliation{Instituto de Física de Partículas y del Cosmos IPARCOS, Facultad de CC. Físicas, Universidad Complutense de Madrid, E-28040 Madrid, Spain}

\author[0000-0002-1836-0771]{Akshatha Gopinath}
\affiliation{Anton Pannekoek Institute for Astronomy, University of Amsterdam, Science Park 904, 1098~XH Amsterdam, the Netherlands}

\author[0000-0003-2317-1446]{Jason~W.~T. Hessels}
\affiliation{Anton Pannekoek Institute for Astronomy, University of Amsterdam, Science Park 904, 1098~XH Amsterdam, the Netherlands}
\affiliation{ASTRON, Netherlands Institute for Radio Astronomy, Oude Hoogeveensedijk 4, 7991~PD Dwingeloo, the Netherlands}
\affiliation{Trottier Space Institute, McGill University, 3550 rue University, Montr\'eal, QC H3A 2A7, Canada}
\affiliation{Department of Physics, McGill University, 3600 rue University, Montr\'eal, QC H3A 2T8, Canada}

\author[0000-0001-6664-8668]{Franz Kirsten}
\affiliation{Department of Space, Earth and Environment, Chalmers University of Technology, Onsala Space Observatory, 439 92, Onsala, Sweden}

\author[0000-0001-9814-2354]{Benito Marcote}
\affiliation{Joint Institute for VLBI ERIC, Oude Hoogeveensedijk 4, 7991~PD Dwingeloo, the Netherlands}
\affiliation{ASTRON, Netherlands Institute for Radio Astronomy, Oude Hoogeveensedijk 4, 7991~PD Dwingeloo, the Netherlands}

\author[0000-0003-3655-2280]{Vladislavs Bezrukovs}
\affiliation{Engineering Research Institute Ventspils International Radio Astronomy Centre (ERI VIRAC) of Ventspils University of Applied Sciences, Inzenieru street 101, Ventspils, LV-3601, Latvia}

\author[0000-0003-1771-1012]{Richard Blaauw}
\affiliation{ASTRON, Netherlands Institute for Radio Astronomy, Oude Hoogeveensedijk 4, 7991~PD Dwingeloo, the Netherlands}

\author[0000-0002-0963-0223]{Justin~D. Bray}
\affiliation{Jodrell Bank Centre for Astrophysics, School of Physics and Astronomy, The University of Manchester, Alan Turing Building, Oxford Road, Manchester, M13 9PL, UK}

\author[0000-0002-3341-466X]{Salvatore Buttaccio}
\affiliation{INAF-Osservatorio Astrofisico di Catania, via Santa Sofia 78, I-95123, Catania, Italy}
\affiliation{INAF-Istituto di Radioastronomia, Via Gobetti 101, 40129, Bologna, Italy}

\author[0000-0003-2047-5276]{Tomas Cassanelli}
\affiliation{Department of Electrical Engineering, Universidad de Chile, Av. Tupper 2007, Santiago 8370451, Chile}

\author[0000-0002-3426-7606]{Pragya Chawla}
\affiliation{ASTRON, Netherlands Institute for Radio Astronomy, Oude Hoogeveensedijk 4, 7991~PD Dwingeloo, the Netherlands}
\affiliation{Anton Pannekoek Institute for Astronomy, University of Amsterdam, Science Park 904, 1098~XH Amsterdam, the Netherlands}

\author[0000-0002-5924-3141]{Alessandro Corongiu}
\affiliation{INAF-Osservatorio Astrofisico di Catania, via Santa Sofia 78, I-95123, Catania, Italy}

\author[0009-0002-1304-0346]{William Deng}
\affiliation{University of Toronto, David Dunlap Institute for Astronomy and Astrophysics, St George St, Toronto, ON M5S 3H4}
\affiliation{University of Cambridge, Institute of Astronomy, Madingley Road, Cambridge CB3 0HA, UK}

\author{Hannah~N. Didehbani}
\affiliation{Department of Physics, Massachusetts Institute of Technology, 77 Massachusetts Ave, Cambridge, MA 02139, USA}

\author[0000-0002-9363-8606]{Yuxin Dong}
\affiliation{Center for Interdisciplinary Exploration and Research in Astrophysics (CIERA) and Department of Physics and Astronomy, Northwestern University, Evanston, IL 60208, USA}

\author[0000-0003-4056-4903]{Marcin~P. Gawro\'nski}
\affiliation{Institute of Astronomy, Faculty of Physics, Astronomy and Informatics, Nicolaus Copernicus University, Grudziadzka 5, PL-87-100 Toru\'n, Poland}

\author[0000-0002-8657-8852]{Marcello Giroletti }
\affiliation{INAF-Istituto di Radioastronomia, Via Gobetti 101, 40129, Bologna, Italy}

\author[0000-0002-9049-8716]{Lucas Guillemot }
\affiliation{Laboratoire de Physique et Chimie de l'Environnement et de l'Espace, Universit\'e d’Orl\'eans/CNRS, 45071 Orl\'eans Cedex 02, France}
\affiliation{Observatoire Radioastronomique de Nan\c{c}ay, Observatoire de Paris, Universit\'e PSL, Universit\'e d’Orléans, CNRS, 18330 Nan\c{c}ay, France}

\author[0000-0002-8043-0048]{Jeff Huang}
\affiliation{Department of Physics, McGill University, 3600 rue University, Montr\'eal, QC H3A 2T8, Canada}
\affiliation{Trottier Space Institute, McGill University, 3550 rue University, Montr\'eal, QC H3A 2A7, Canada}

\author{Dmitriy~V. Ivanov}
\affiliation{Institute of Applied Astronomy, Kutuzova Embankment 10, St. Petersburg, 191187, Russia}

\author[0000-0003-3457-4670]{Ronniy~C. Joseph}
\affiliation{Department of Physics, McGill University, 3600 rue University, Montr\'eal, QC H3A 2T8, Canada}
\affiliation{Trottier Space Institute, McGill University, 3550 rue University, Montr\'eal, QC H3A 2A7, Canada}

\author[0000-0001-9345-0307]{Victoria~M. Kaspi}
\affiliation{Department of Physics, McGill University, 3600 rue University, Montr\'eal, QC H3A 2T8, Canada}
\affiliation{Trottier Space Institute, McGill University, 3550 rue University, Montr\'eal, QC H3A 2A7, Canada}

\author[0000-0002-0321-8588]{Mikhail~A. Kharinov}
\affiliation{Institute of Applied Astronomy, Kutuzova Embankment 10, St. Petersburg, 191187, Russia}

\author[0000-0002-5857-4264]{Mattias Lazda}
\affiliation{University of Toronto, David Dunlap Institute for Astronomy and Astrophysics, St George St, Toronto, ON M5S 3H4}
\affiliation{David A.~Dunlap Department of Astronomy \& Astrophysics, University of Toronto, 50 St.~George Street, Toronto, ON M5S 3H4, Canada}

\author[0000-0002-3669-0715]{Michael Lindqvist}
\affiliation{Department of Space, Earth and Environment, Chalmers University of Technology, Onsala Space Observatory, 439 92, Onsala, Sweden}

\author[0000-0002-1482-708X]{Maccaferri	Giuseppe}
\affiliation{INAF-Istituto di Radioastronomia, Via Gobetti 101, 40129, Bologna, Italy}

\author[0000-0003-4584-8841]{Lluis Mas-Ribas}
\affiliation{University of California, Santa Cruz, UCO Observatories, 1156 High St., Santa Cruz, CA 95064, USA}

\author[0000-0002-4279-6946]{Kiyoshi~W. Masui}
\affiliation{MIT Kavli Institute for Astrophysics and Space Research, Massachusetts Institute of Technology, 77 Massachusetts Ave, Cambridge, MA 02139, USA}
\affiliation{Department of Physics, Massachusetts Institute of Technology, 77 Massachusetts Ave, Cambridge, MA 02139, USA}

\author[0000-0001-7348-6900]{Ryan Mckinven}
\affiliation{Department of Physics, McGill University, 3600 rue University, Montr\'eal, QC H3A 2T8, Canada}
\affiliation{Trottier Space Institute, McGill University, 3550 rue University, Montr\'eal, QC H3A 2A7, Canada}

\author[0000-0002-8466-7026]{Alexey Melnikov}
\affiliation{Institute of Applied Astronomy, Kutuzova Embankment 10, St. Petersburg, 191187, Russia}

\author[0000-0002-2551-7554]{Daniele Michilli}
\affiliation{MIT Kavli Institute for Astrophysics and Space Research, Massachusetts Institute of Technology, 77 Massachusetts Ave, Cambridge, MA 02139, USA}
\affiliation{Department of Physics, Massachusetts Institute of Technology, 77 Massachusetts Ave, Cambridge, MA 02139, USA}

\author[0000-0002-3355-2261]{Andrey~G. Mikhailov}
\affiliation{Institute of Applied Astronomy, Kutuzova Embankment 10, St. Petersburg, 191187, Russia}

\author[0000-0002-2028-9329]{Anya~E. Nugent}
\affiliation{Center for Interdisciplinary Exploration and Research in Astrophysics (CIERA) and Department of Physics and Astronomy, Northwestern University, Evanston, IL 60208, USA}

\author[0000-0001-9381-8466]{Omar~S. Ould-Boukattine}
\affiliation{ASTRON, Netherlands Institute for Radio Astronomy, Oude Hoogeveensedijk 4, 7991~PD Dwingeloo, the Netherlands}
\affiliation{Anton Pannekoek Institute for Astronomy, University of Amsterdam, Science Park 904, 1098~XH Amsterdam, the Netherlands}

\author[0000-0002-5195-335X]{Zsolt Paragi}
\affiliation{Joint Institute for VLBI ERIC, Oude Hoogeveensedijk 4, 7991~PD Dwingeloo, the Netherlands}

\author[0000-0002-8912-0732]{Aaron~B. Pearlman}
\affiliation{Department of Physics, McGill University, 3600 rue University, Montr\'eal, QC H3A 2T8, Canada}
\affiliation{Trottier Space Institute, McGill University, 3550 rue University, Montr\'eal, QC H3A 2A7, Canada}
\affiliation{Banting Fellow}
\affiliation{McGill Space Institute Fellow}
\affiliation{FRQNT Postdoctoral Fellow}

\author[0000-0003-2155-9578]{Ue-Li Pen}
\affiliation{Canadian Institute for Theoretical Astrophysics, University of Toronto, Ontario, Canada}
\affiliation{David A.~Dunlap Department of Astronomy \& Astrophysics, University of Toronto, 50 St.~George Street, Toronto, ON M5S 3H4, Canada}

\author[0000-0002-4795-697X]{Ziggy Pleunis}
\affiliation{Anton Pannekoek Institute for Astronomy, University of Amsterdam, Science Park 904, 1098~XH Amsterdam, the Netherlands}
\affiliation{ASTRON, Netherlands Institute for Radio Astronomy, Oude Hoogeveensedijk 4, 7991~PD Dwingeloo, the Netherlands}

\author[0000-0003-3154-3676]{Ketan~R. Sand}
\affiliation{Department of Physics, McGill University, 3600 rue University, Montr\'eal, QC H3A 2T8, Canada}
\affiliation{Trottier Space Institute, McGill University, 3550 rue University, Montr\'eal, QC H3A 2A7, Canada}

\author[0000-0002-4823-1946]{Vishwangi Shah}
\affiliation{Department of Physics, McGill University, 3600 rue University, Montr\'eal, QC H3A 2T8, Canada}
\affiliation{Trottier Space Institute, McGill University, 3550 rue University, Montr\'eal, QC H3A 2A7, Canada}

\author[0000-0002-6823-2073]{Kaitlyn Shin}
\affiliation{MIT Kavli Institute for Astrophysics and Space Research, Massachusetts Institute of Technology, 77 Massachusetts Ave, Cambridge, MA 02139, USA}
\affiliation{Department of Physics, Massachusetts Institute of Technology, 77 Massachusetts Ave, Cambridge, MA 02139, USA}

\author[0000-0001-6170-2282]{Mark P. Snelders}
\affiliation{Anton Pannekoek Institute for Astronomy, University of Amsterdam, Science Park 904, 1098~XH Amsterdam, the Netherlands}
\affiliation{ASTRON, Netherlands Institute for Radio Astronomy, Oude Hoogeveensedijk 4, 7991~PD Dwingeloo, the Netherlands}

\author[0000-0002-8476-6307]{Tiziana Venturi}
\affiliation{INAF-Istituto di Radioastronomia, Via Gobetti 101, 40129, Bologna, Italy}

\author[0000-0002-9786-8548]{Na Wang}
\affiliation{Xinjiang Astronomical Observatory, CAS, 150 Science 1-Street, Urumqi, Xinjiang 830011, China}

\author[0000-0001-7361-0246]{David~R.~A. Williams-Baldwin}
\affiliation{Jodrell Bank Centre for Astrophysics, School of Physics and Astronomy, The University of Manchester, Alan Turing Building, Oxford Road, Manchester, M13 9PL, UK}

\author[0000-0002-2322-5232]{Jun Yang}
\affiliation{Xinjiang Astronomical Observatory, CAS, 150 Science 1-Street, Urumqi, Xinjiang 830011, China}

\author{Jianping~P. Yuan}
\affiliation{Xinjiang Astronomical Observatory, CAS, 150 Science 1-Street, Urumqi, Xinjiang 830011, China}

\correspondingauthor{Jason W.~T. Hessels}
\email{jason.hessels@mcgill.ca}



\begin{abstract}
We present the localization and host galaxy of \rtwelve, a repeating source of fast radio bursts (FRBs) discovered using CHIME/FRB. As part of the PRECISE repeater localization program on the EVN, we monitored \rtwelve for 65.6\,hours at $\sim1.4$\,GHz and detected a single burst, which led to its VLBI localization with 260\,mas uncertainty (2$\sigma$). Follow-up optical observations with the MMT Observatory ($i\gtrsim 25.7$\,mag (AB)) found no visible host at the FRB position. Subsequent deeper observations with the GTC, however, revealed an extremely faint galaxy ($r=27.32 \pm0.16$\,mag), very likely ($99.95\,\%$) associated with \rtwelve. Given the dispersion measure of the FRB ($\sim580$\,pc\,cm$^{-3}$), even the most conservative redshift estimate ($z_{\mathrm{max}}\sim0.83$) implies that this is the lowest-luminosity FRB host to date ($\lesssim10^8L_{\odot}$), even less luminous than the dwarf host of FRB~20121102A. We investigate how localization precision and the depth of optical imaging affect host association, and discuss the implications of such a low-luminosity dwarf galaxy. Unlike the other repeaters with low-luminosity hosts, \rtwelve has a modest Faraday rotation measure of a few tens of rad\,m$^{-2}$, and EVN plus VLA observations reveal no associated compact persistent radio source. We also monitored \rtwelve for 40.4\,hours over 2\,years as part of the \'ECLAT repeating FRB monitoring campaign on the Nan\c{c}ay Radio Telescope, and detected one burst. Our results demonstrate that, in some cases, the robust association of an FRB with a host galaxy will require both high localization precision, as well as deep optical follow-up.
\end{abstract}



\section{Introduction} \label{sec:intro}

Fast radio bursts (FRBs) are a class of extremely luminous, extragalactic, coherent radio transients that have durations on the order of milliseconds or less \citep[for a review see, e.g.,][]{petroff_2022_aarv}. The majority of FRBs are observed as single events, but a small fraction \citep[$\sim$3\,\%;][]{chime_2023_apj} are known to be  repeaters \citep{spitler_2016_natur}, from which multiple bursts have been detected. The repetition rate varies significantly among repeaters. While most repeaters have low repetition rates ($\sim10^{-3}$--$10^{-1}$\,bursts\,hr$^{-1}$ above a fluence threshold of 5\,Jy\,ms), on par with the upper limits on the repetition rates of apparent non-repeaters \citep{chime_2023_apj}, some boast repetition rates about an order-of-magnitude larger, e.g., \rsixtyseven \citep{lanman_2022_apj} and \rnarwhal \citep{mckinven_2022_atel}. Broadband studies of a few active repeaters suggests that these rates are likely frequency-dependent \citep[e.g.,][]{josephy_2019_apjl,chawla_2020_apjl}, but most of the less active CHIME/FRB repeaters have not been well-studied at higher frequencies. 

The emission mechanism(s) and progenitor(s) of FRBs are not fully understood, but their short durations and high brightness temperatures ($\sim10^{37}$\,K) suggest neutron star or black hole origins, with magnetars being strong candidates in particular, given the energy demands of some high-repetition sources \citep[e.g.,][]{metzger_2017_apj}. The magnetar hypothesis was further reinforced by the detection of an FRB-like burst from the Galactic magnetar, SGR~1935+2154 \citep[][]{bochenek_2020_natur,chime_2020_natur_galacticfrb}. 

Identifying the host galaxies of FRBs requires roughly arcsecond precision (or better) on the localization of the radio bursts \citep{eftekhari_2017_apj}. Studies of FRB hosts find that the majority are consistent with neutron star progenitors produced in core-collapse supernovae, but a small fraction are more consistent with progenitors from older stellar populations \citep{gordon_2023_apj,2024ApJ...971L..51B,law_2024_apj}. While there is no clear statistical distinction between the hosts of repeaters and non-repeaters \citep{2024ApJ...971L..51B,gordon_2023_apj}, tentatively, the hosts of repeaters do extend to lower masses, while non-repeater hosts tend to be more optically luminous \citep{gordon_2023_apj,heintz_2020_apj}.

Of the 45 FRBs that have been associated with a host galaxy\footnote{The FRB Community Newsletter (Volume 05, Issue 07, DOI: 10.7298/PRE0-VF51).}, seven repeaters have been localized to milliarcsecond precision using the European VLBI Network (EVN), thus enabling characterization of the parsec-scale environment surrounding the source \citep{marcote_2017_apjl,marcote_2020_natur,kirsten_2022_natur,nimmo_2022_apjl,bhandari_2023_apjl,hewitt_2024_mnras,snelders_2024_atel}. Together with the challenges of acquiring very-long-baseline interferometry (VLBI) observations of sporadic transients, multi-wavelength observations with matching resolution are required to leverage the astrometric precision achieved for the bursts themselves. These localizations revealed that some active repeaters are associated with star-forming regions -- ideal birth places of magnetars formed via the core-collapse of massive stars \citep{chittidi_2021_apj,dong_2024_apj} -- while some are slightly offset from local peaks of star formation \citep{bassa_2017_apjl,tendulkar_2021_apjl}. One repeater even inhabits a globular cluster \citep{kirsten_2022_natur}, necessitating delayed magnetar formation channels, such as binary neutron star merger or accretion-induced collapse of a white dwarf, for at least some sources \citep{kremer_2021_apjl}.

The focus of this present work is on \rtwelve, a repeater discovered by CHIME/FRB \citep{fonseca_2020_apjl}. To date, CHIME/FRB has detected 15 bursts in total from this source\footnote{\url{www.chime-frb.ca/repeaters/FRB20190208A}}, including seven baseband events (with raw-voltage data recorded) between 2020 January and 2021 December \citep{mckinven_2023_apj}. The bursts show high fractions of linear polarization and some tentative evidence of minor fluctuations in the polarization position angle (PPA) towards the edges of the bursts. The reported dispersion measure (DM) has been stable around 580\,pc\,cm$^{-3}$ across 2020 and 2021, while the observed Faraday rotation measure (RM) has shown `u-shaped' evolution over this same period: decreasing from $\sim$30\,rad\,m$^{-2}$ to $\sim$10\,rad\,m$^{-2}$, before increasing again to $\sim$30\,rad\,m$^{-2}$. To our knowledge there are no other detections of \rtwelve in the literature.

In this paper, we present a VLBI localization of \rtwelve using the EVN. Leveraging this precise localization, we performed optical follow-up observations to identify the host galaxy and searched for any associated compact persistent radio emission. We also present the results of two targeted radio monitoring campaigns at $\sim1.4$\,GHz, which totaled more than 100 hours of exposure, and resulted in two bursts being detected from \rtwelve. These are the first bursts that have been observed from this source by a telescope other than CHIME/FRB. Section~\ref{sec:radobs} describes the various radio observations, the source localization, and the properties of the detected bursts. Section~\ref{sec:optobs} describes the subsequent optical observations and host galaxy association. In Section~\ref{sec:disc}, we discuss our results. Detailed descriptions of pipelines, observation logs and analyses of the bursts are presented in Appendices~\ref{app:evn} to \ref{app:burpop}. Throughout this work we assume Planck18 cosmological parameters \citep{planckcollaboration_2020_aa}.

\section{Radio observations and interferometric localization}
\label{sec:radobs}

\subsection{EVN Observations}
We observed the \rtwelve field 38 times between 2021 February and 2023 August using an {\it ad-hoc} array of EVN dishes, in `EVN-Lite' mode\footnote{EVN-Lite is a new initiative to address rare/transient phenomena requiring hundreds of hours of observing time with ad hoc
subarrays of radio telescopes that form the EVN, outside the regular EVN
observing sessions.}, at a central frequency of $\sim1.4$\,GHz. The total exposure time on \rtwelve, accounting for phase-referencing scans of a nearby calibrator, was 65.6\,hours. These observations were carried out as part of the ongoing FRB VLBI localization campaign called PRECISE \citep[Pinpointing REpeating ChIme Sources with EVN dishes; PI: F.~Kirsten; see, e.g.,][]{marcote_2022_evlb}. We detected a single burst (which we will refer to as B1) in one observation, described below. The other observations had a similar setup in terms of calibrator source selection, observing strategy and recording details, but the array configuration somewhat differed depending on the availability of dishes. In all observations, the 100-m Effelsberg telescope participated, and these single-dish data were searched for bursts. 

\begin{figure*}
    \centering
    \includegraphics[width=.475\textwidth]{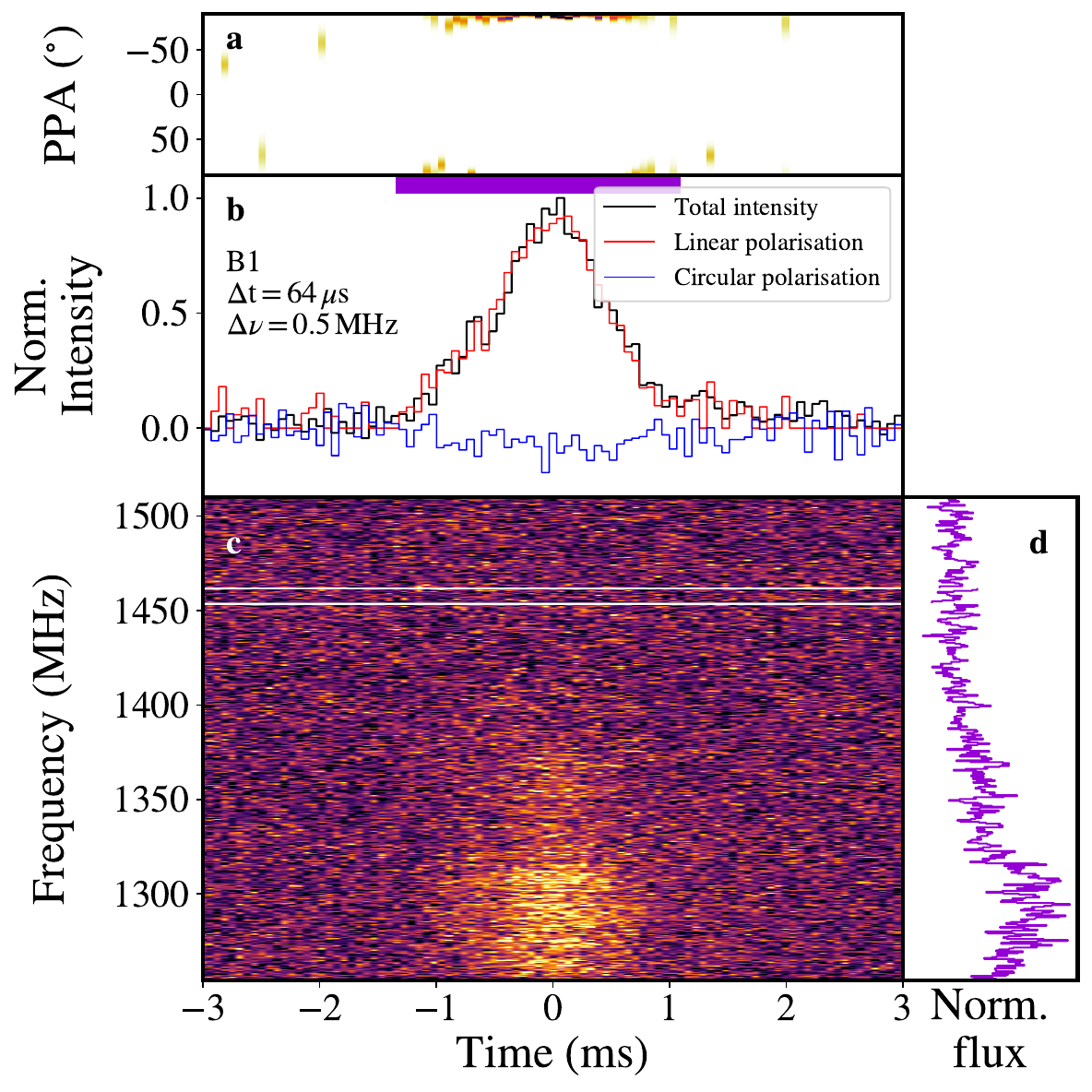}
    \includegraphics[width=.475\textwidth]{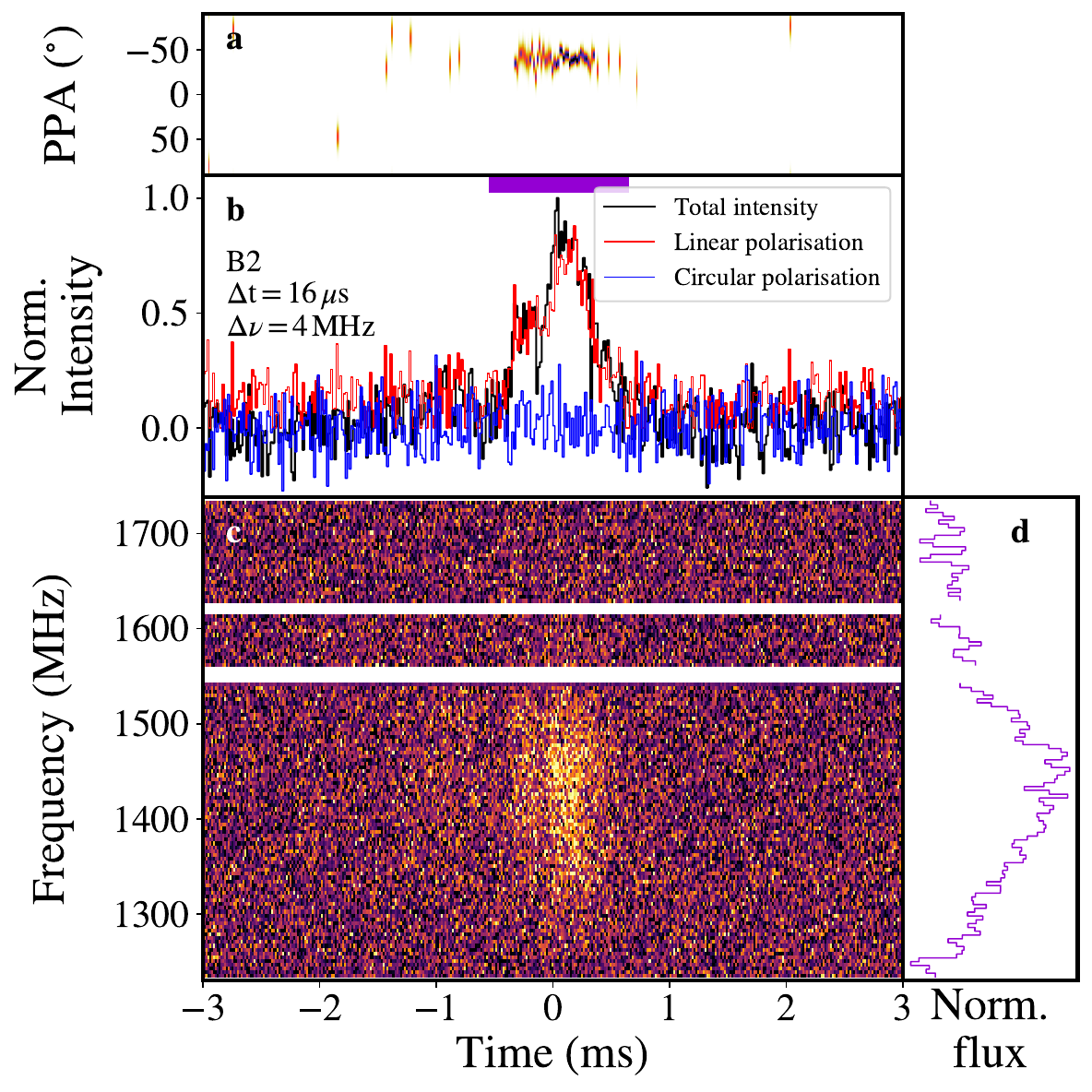}
    \caption{The left and right panels show bursts B1 and B2, detected in the PRECISE and \'ECLAT projects, respectively. In each sub-figure, a dynamic spectrum of the burst is shown in Panel~c. Bursts B1 and B2 have been dedispersed using DMs of 580.1 and 579.9\,pc\,cm$^{-3}$, respectively. The horizontal white lines indicate channels contaminated by RFI that have been masked. A frequency-averaged (over the entire observing band) total intensity profile of the burst is shown in black in Panel~b. The unbiased linear polarization is shown in red and the circular polarization in blue. In Panel~a, the probability distribution function of the PPA, at each time step, is shown. The bursts have been derotated according to their respective RMs (46.5\,rad\,m$^{-2}$ for B1 and $-5.2$\,rad\,m$^{-2}$ for B2), and the PPA is referenced to infinite frequency. In Panel~d, the time-averaged frequency spectrum (integrated over the burst duration indicated by the horizontal purple bar in Panel~b) is shown in purple. Note that the observing bandwidths differ between the two instruments.}
    \label{fig:portraits}
\end{figure*}

We detected B1 (left panel in Figure~\ref{fig:portraits}) in an observation on 2021 October 17, lasting from 08:49~UT to 11:23~UT (PRECISE project code PR187A), using nine EVN dishes: Badary, Effelsberg, Irbene, Medicina, Onsala, Svetloe, Toruń, Urumqi, and Zelenchukskaya. Detailed descriptions of the observational setup and search pipeline are presented in Appendix~\ref{app:evn}.

We observed a test pulsar,  PSR~B2255+58, for 5\,minutes at the beginning and end of the observation to assess data quality and verify the polarimetric calibration. After the first test pulsar scan, we observed J1419+5423 for 5\,minutes to use as a fringe finder and bandpass calibrator. For phase referencing, we alternated between 1.5-minute scans of our chosen phase calibrator, J1852+4855 (2.0\,degrees offset from the preliminary \rtwelve position\footnote{An arcminute-level position for \rtwelve was made available through a Memorandum of Understanding between the PRECISE project and the CHIME/FRB Collaboration.}) and 5.5-minute scans of \rtwelve. Finally, we also observed the source J1850+4959 (1.1\,degrees offset from the phase calibrator) as an interferometric check source (see Section~\ref{sec:loc} and Appendix~\ref{app:evn}). The total exposure time on \rtwelve during this one session with a detected burst was $\sim97$\,minutes.

\subsection{NRT Observations}
The \'ECLAT (Extragalactic Coherent Light from Astrophysical Transients; PI: D.~Hewitt) observing campaign on the Nan\c{c}ay Radio Telescope (NRT), has been performing targeted follow-up observations of repeating FRBs since the start of 2022. About 20 CHIME/FRB repeaters are observed for approximately one hour per week each. Observations are conducted at a central frequency of 1.484\,GHz using the low-frequency receiver ($1.1-1.8$\,GHz) of the focal plane and receivers system, FORT (\textit{Foyer Optimis\'e pour le Radio T\'elescope}). The Nan\c{c}ay Ultimate Pulsar Processing Instrument \citep[NUPPI;][]{desvignes_2011_aipc} records full-polarization data (in a linear basis) with 32-bit sampling, 16\,$\upmu$s time resolution, and a total observing bandwidth of 512\,MHz -- consisting of eight 64-MHz subbands, divided into 4-MHz channels. We applied coherent dedispersion (i.e., dedispersion \textit{within} spectral channels) at a DM of 579\,pc\,cm$^{-3}$ for \rtwelve\footnote{This is the rounded DM of the most recent \rtwelve burst detected by CHIME/FRB at the start of \'ECLAT monitoring.}. Additionally, a 10-s observation of a 3.33-Hz pulsed noise diode is also acquired with each \'ECLAT FRB observation for polarimetric calibration.

Between 2022 February and 2023 December, \rtwelve was observed 52 times, resulting in a total exposure time of 40.4\,hours. The search pipeline is summarized in Appendix~\ref{app:nrt}.  We detected a single \rtwelve burst in our \'ECLAT observations on 2023 May 17, which we will refer to as B2 (right panel in Figure~\ref{fig:portraits}).

\subsection{VLA Observations}
We searched for persistent radio continuum emission in the field of \rtwelve using the Karl G. Jansky Very Large Array (VLA) in C-configuration. Observations were conducted on 2021 October 19 23:36:00~UT as part of program VLA/22B-126 (PI: S.~Bhandari) in the $4\text{--}8$\,GHz C-band, divided into $32 \times 128$\,MHz spectral windows and centred at a frequency of 6\,GHz. Sources 3C286 and J1852$+$4855 were used as flux and phase calibrators, respectively.  
The target field was observed for 71.6\,min yielding an RMS noise of 4\,$\upmu$Jy\,beam$^{-1}$, but no persistent radio source was detected at the position of \rtwelve. We used the VLA pipeline calibration and performed imaging, with Briggs' weighting (robust=0.5) and a cellsize of $1^{\prime\prime}$, in \texttt{CASA} \citep{mcmullin_2007_aspc,vanbemmel_2022_pasp} using the task \texttt{tclean}. The resulting synthesized beam size was $3.2^{\prime\prime}\times2.8^{\prime\prime}$.

\subsection{Properties of the Bursts}
\label{sec:burst_prop}

\begin{table*} 
    \caption{Properties of the bursts detected from \rtwelve at 1.4\,GHz}
    \label{tab:properties}
    \begingroup
    \centering
    \begin{tabular}{lcc} \hline
        Property                &   B1  & B2 \\ \hline
        Telescope               &   EVN     & NRT \\
        TOA$^a$                 &   59504.40346933  & 60081.13824472   \\
        Burst width$^b$ (ms)    &   1.13 $\pm$ 0.03   & 0.65 $\pm$ 0.03    \\
        Fluence$^c$ (Jy\,ms)    &   1.70 $\pm$ 0.34            & 0.95 $\pm$ 0.19            \\
        Bandwidth$^d$ (MHz)         &   $>$ 97 $\pm$ 5        & 191 $\pm$ 8        \\
        DM$_{\rm S/N}$$^e$ (pc\,cm$^{-3}$)  & 580.01 $\pm$ 0.26 & 580.03 $\pm$ 0.14 \\
        DM$_{\rm struct}$$^f$ (pc\,cm$^{-3}$)  &  580.24 $\pm$ 0.23 & 579.84 $\pm$ 0.29 \\
        RM$_{\rm obs}$$^g$ (rad\,m$^{-2}$)  & +46.5 $\pm$ 16.5  & -5.2 $\pm$ 4.9 \\
        RM$_{\rm iono}$$^h$ (rad\,m$^{-2}$)  & 1.07 $\pm$ 0.07   &   1.39 $\pm$ 0.09 \\
         \hline
    \end{tabular}
    \\
    \endgroup
    $^a$  The burst time-of-arrival at the Solar system barycentre in TDB, corrected to infinite frequency for a DM of 580\,pc\,cm$^{-3}$ and using a DM constant of 1/(2.41$\times$10$^{-4}$ ) MHz$^2$\,pc$^{-1}$ \,cm$^{3}$\,s. Measured at the time bin corresponding to the peak flux density. The position for Effelsberg is X = 4033947.2355\,m, Y = 486990.7943\,m, Z = 4900431.0017\,m , and for the NRT X = 4324165.81 m, Y = 165927.11\,m, Z = 4670132.83\,m.\\
    $^b$FWHM of a Gaussian fit to the frequency-averaged profile.\\
    $^c$ We assume an uncertainty of approximately 20\,\%, dominated by the uncertainty on the system equivalent flux density (SEFD). \\
    $^d$ FWHM of a Gaussian fit to the time-averaged spectrum.\\
    $^e$ DM determined from S/N optimisation (see Appendix~\ref{sec:dm}).\\
    $^f$ DM determined from structure optimisation using \texttt{DM\_phase} \citep[][ see Section~\ref{sec:dm}]{seymour_2019_ascl}.\\
    $^g$ See Section~\ref{sec:pol_eff} and \ref{sec:pol_nrt}. We caution that there is a sign ambiguity. \\
    $^h$ The expected ionospheric RM contributions using IonFR \citep{sotomayorbeltran_2013_ascl}.\\
\end{table*}

The properties of bursts B1 and B2 are summarized in Table~\ref{tab:properties}. In this section we briefly summarize these properties, while detailed description of the analyses are presented in Appendix~\ref{app:burpop}. 

We estimate the DM of B1 and B2 to be 580.1 and 579.9\,pc\,cm$^{-3}$, respectively, which is consistent with previous estimates by CHIME/FRB \citep{fonseca_2020_apjl,mckinven_2023_apj}. The estimated DM contribution from the Milky Way thin/thick-disk ISM is 71.5\,pc\,cm$^{-3}$ \citep[using NE2001p][]{ocker_2024_rnaas}. The expected scattering timescales from the Milky Way at the center frequencies of the PRECISE and \'ECLAT observing bands are 0.137 and 0.103\,$\upmu$s, respectively, significantly smaller than the time resolution of our data. The lack of substantial scattering is also visible in the burst profiles (Figure~\ref{fig:portraits}), and bodes well for probing the bursts for microstructure. To do so, we produced filterbank data for B1 at time resolutions of 64, 16 and 4\,$\upmu$s, but find no prominent structure on these time scales. We find RMs of $+46.5\pm16.5$ and $-5.2\pm4.9$\,rad\,m$^{-2}$ for B1 and B2, respectively. These values have not been corrected for the Galactic RM contribution for this line-of-sight \citep[$+4\pm12$\,rad\,m$^{-2}$][]{hutschenreuter_2022_aa}, or the expected ionospheric contributions of $1.07\pm0.07$ and $1.39\pm0.09$\,rad\,m$^{-2}$ for B1 and B2, respectively. Taking uncertainties into account, the RM for B1 is consistent with RM$_{\text{QU}}=25.75\pm18$\,rad\,m$^{-2}$ measured for a CHIME/FRB \rtwelve burst, detected eight days earlier \citep{mckinven_2023_apj}. All together, these burst properties and propagation effects indicate a relatively clean line-of-sight towards the source. 

\rtwelve appears to be more active at CHIME/FRB frequencies than at $\sim$1.4\,GHz. To quantify the frequency-dependent activity, we calculate the statistical spectral index, which compares the rates at two different frequencies, taking into account instrumental sensitivity. We find values of $\alpha_{\rm{s,NRT/CHIME}}=-2.30\pm0.46$ and $\alpha_{\rm{s,EFF/CHIME}}=-2.96\pm0.49$. Within errors, these values are more or less comparable to what has been measured for other repeaters like \rone \citep{houben_2019_aa} and \rthree \citep{chawla_2020_apjl}. Notably, all these sources appear to be less active at higher frequencies.

\subsection{EVN Correlation and Localization}
\label{sec:loc}
A detailed description of correlation passes, the step-by-step localization procedure and the FRB positional uncertainty estimation is presented in Appendix~\ref{app:evn}.

The PRECISE data underwent multiple correlation iterations at the Joint
Institute for VLBI ERIC (JIVE) in the Netherlands (EVN correlation proposal EK050; PI: F.~Kirsten) using the software correlator \texttt{SFXC} \citep{keimpema_2015_exa}. 

We calibrated and imaged the EVN data using standard interferometric techniques in \texttt{AIPS} \citep{greisen_2003_assl} and \texttt{DIFMAP} \citep{shepherd_1994_baas}. Calibration solutions were derived from the fringe finder (J1419+5423) and phase calibrator (J1852+4855, with a positional uncertainty of $\Delta\alpha=0.15\,$mas, $\Delta\delta=0.10\,$mas from the RFC 2023B catalogue\footnote{\url{astrogeo.org/sol/rfc/rfc_2023b/rfc_2023b_cat.html}}), before being applied to the interferometric check source (J1850+4959) to verify that the calibration was successful, and to help assess the precision of our localization. Using \texttt{DIFMAP}, with a natural weighting scheme and cell size of 1\,mas (in each dimension), we imaged the check source and measured its position. Compared to its position in the RFC 2023B catalogue (which has an uncertainty of $\Delta\alpha=0.20\,$mas, $\Delta\delta=0.15\,$mas), we found a positional offset of $\Delta\alpha=1.46\,$mas, $\Delta\delta=0.03\,$mas. Given our synthesized beam size of $54\,\times23$\,mas, we thus conclude that our calibration was successful. The calibration solutions were then applied to the target field of \rtwelve before imaging (again using \texttt{DIFMAP} with natural weighting and a cell size of 1\,mas). 

\begin{figure}
    \centering
    \includegraphics[width=0.5\textwidth]{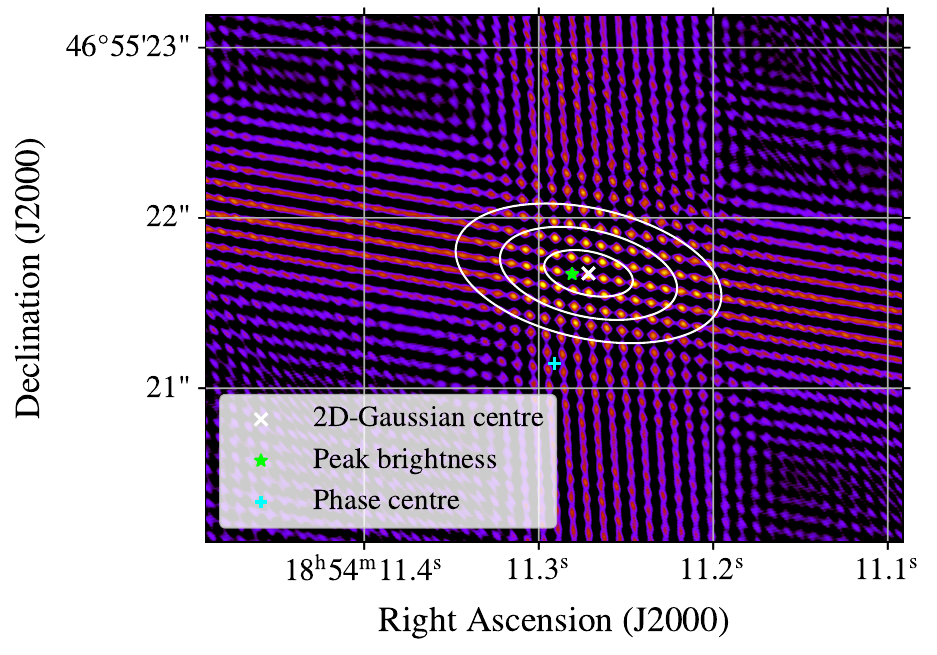}
    \caption{Dirty map of B1, the burst from \rtwelve detected in PRECISE observations. The cyan plus sign indicates the position of the phase center; the green star shows the position of the pixel with the peak brightness; and the white cross the marks center of the Gaussian ellipse.}
    \label{fig:localisation}
\end{figure}

Unfortunately, the spectral extent of B1 was covered by the observing bands of only three dishes: Effelsberg, Onsala and Toruń (Figure~\ref{fig:freq_coverage}). Fortunately, however, the two baselines given by these dishes are oriented nearly orthogonal with respect to each other, and so the fringe pattern forms a cross-like shape, shown in Figure~\ref{fig:localisation}. Although the subarcsecond position of \rtwelve is clear from this map, the relatively poor uv-coverage results in less precise localization compared to previous EVN VLBI localizations of repeaters, though still on par with what is expected from localizations of individual bursts \citep[see e.g.,][]{nimmo_2022_apjl,hewitt_2024_mnras}. 

In the case of \rtwelve presented here, there are multiple sidelobes of comparable brightness, where the fringes from the baselines intersect. Providing a  statistical confidence region for the FRB position is thus non-trivial. To quantify the localization region, we fit a 2D Gaussian function to all pixels where the absolute value of the dirty map of the burst is above 3$\sigma$. This 2D Gaussian fit has $\sigma_{\rm x}=127$\,mas, $\sigma_{\rm y}=265$\,mas and $\theta=0.217$\,rad (measured clockwise in the dirty map presented in Figure~\ref{fig:localisation}). The ellipses shown in Figure~\ref{fig:localisation} are the 1-,2- and 3-$\sigma$ contours of this Gaussian fit. These are not traditional confidence error ellipses, rather these ellipses contain 68.3, 95.4 and 99.7\,\% of the map pixels with power above 3$\sigma$. The center position of this 2D Gaussian (the white cross in the map), is located near the nominal peak brightness position (the green star). 

The position we find for \rtwelve, with conservative uncertainty estimates from the 2-$\sigma_{\rm y}$ contour of the 2D-Gaussian fit, is: \\
\noindent RA (J2000) = $18^{\rm h}54^{\rm m}11\fs27\pm 260$\,mas \\
Dec (J2000) = $+46^{\circ}55'21\farcs67\pm 260$\,mas.\\

\section{Optical observations and host galaxy association}
\label{sec:optobs}

\begin{figure*}
\centering
\includegraphics[width=0.5\textwidth]{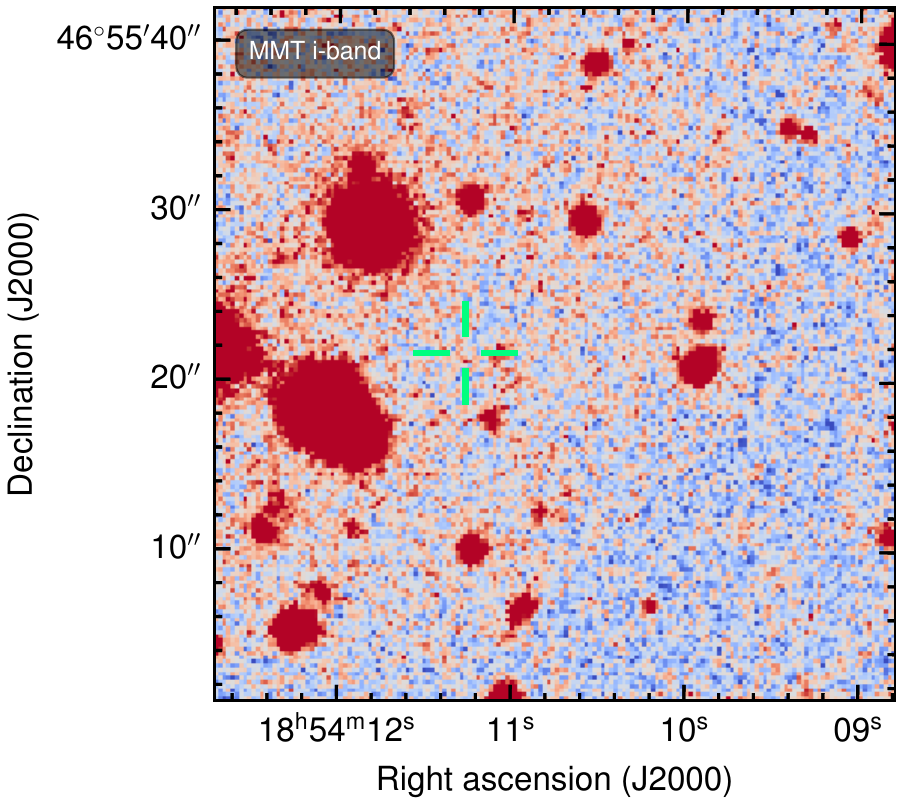}\hfill    
\includegraphics[width=0.5\textwidth]{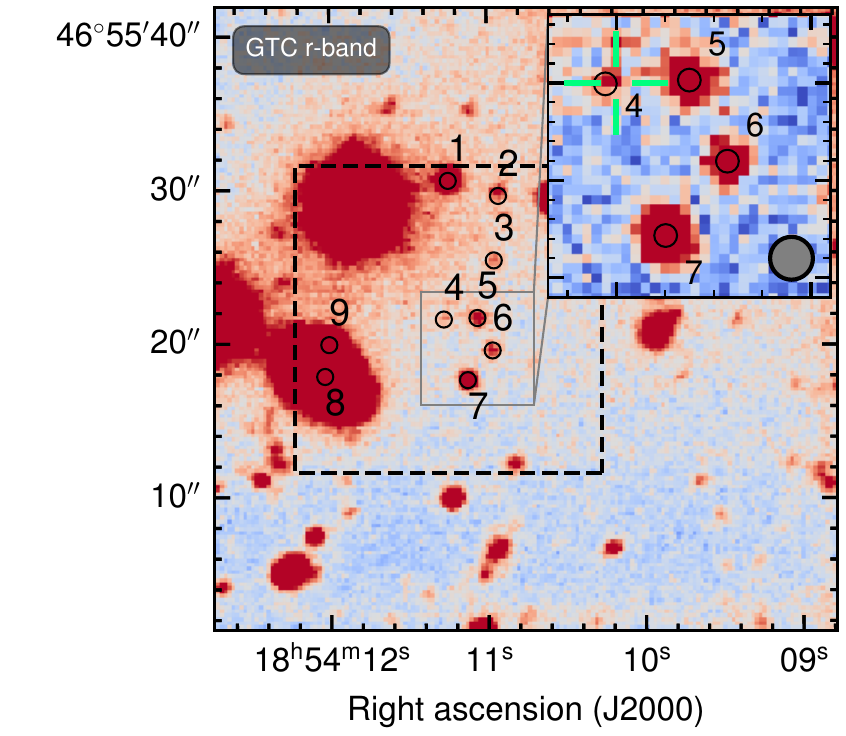}
\caption{
\textbf{Left:} MMT $i$-band image of the field surrounding the \rtwelve localization. No host is visible, with a $3\sigma$ limit of $i\gtrsim 25.7$\,mag (AB) at the position of \rtwelve. 
\textbf{Right:} GTC $r$-band image showing the same field, but achieving about two magnitudes greater depth (see Section \ref{sec:optical_host}). The dashed black box indicates the region that was considered in host association, and the nine sources that were identified as potential candidate hosts are marked by black circles (see Section \ref{sec:host_ass}). The green crosshair indicates the FRB position -- note that the EVN burst localization region is much smaller, roughly the pixel scale in this image. In the GTC image the FRB position is spatially coincident with source~O4 (also see Table~\ref{tab:host}), which is not visible in the MMT image. The gray disk in the bottom-right of the inset indicates the effective point-spread function of the GTC image. The seeing varied in the range $0.6$--$0.9^{\prime\prime}$ during the GTC observations.}
\label{fig:FOV_optical}
\end{figure*}

\label{sec:optical_host}
\subsection{MMT Observations}
We observed the field of \rtwelve\ using the Binospec imaging spectrograph \citep{fabricant_2019_pasp} mounted on the 6.5-m MMT Observatory on 2022 May 31 UT (PI: A.~Nugent; Program 2022A-UAO-G193) in the $i$-band for a total of 2235\,s of exposure. We used the custom \texttt{POTPyRI} pipeline\footnote{\url{https://github.com/CIERA-Transients/POTPyRI}} to apply bias and flat-field corrections, perform cosmic ray rejection and image co-addition, and to perform astrometric calibration to the {\it Gaia} DR3 catalog \citep{gaiacollaboration_2023_aa} on the final co-added image stack. While there are extended sources in the vicinity (Figure~\ref{fig:FOV_optical}), no source was detected at the VLBI-position of \rtwelve (determined in Section~\ref{sec:loc}). To obtain a photometric limit at this position, we performed aperture photometry at the FRB position and the surrounding extended sources, using a custom script based on the \texttt{aperture\char`_photometry} module of \texttt{photutils} \citep{photutils}\footnote{\url{https://github.com/charliekilpatrick/photometry}} and calculated a 3-$\sigma$ limit of $i \gtrsim 25.7$\,mag (AB) at the position of \rtwelve. 

We also obtained $2\times900$\,s of spectroscopy of two sources in the field (O8 and O9; see Section~\ref{sec:host_ass} and Figure~\ref{fig:FOV_optical} for the naming convention) with MMT/Binospec, in which the slit was aligned to capture both sources. O8 was initially considered to be a plausible host, and the nearby O9 was able to be covered in the same slit. We used the 270 lines/mm grating with the LP3800 blocking filter and a central wavelength 6500\,\AA\ to cover a wavelength range of $3850-9150$\,\AA. The data were reduced using the Python Spectroscopic Data Reduction Pipeline (\texttt{PypeIt}; \citealt{pypeit:zenodo,pypeit:joss_pub}) in the \textit{quicklook} reduction mode to identify the redshifts of galaxies O8 and O9. For galaxy O8, we identify eight spectral features at a common redshift of $z=0.1935\pm0.0203$. For galaxy O9, we identify three emission features at a common redshift of $z=0.5473\pm0.0004$.

\subsection{GTC Observations}
The absence of an obvious source at the position of \rtwelve motivated deeper follow-up. Thus, we obtained observations with the Optical System for Imaging and low Resolution Integrated Spectroscopy \citep[OSIRIS;][]{Cepa_2000_SPIE} mounted on the 10.4-m GTC on 2024 June 29 and July 1 UT (PI: A. Gil de Paz; Program GTCMULTIPLE2G-24A) in the Sloan $r$-band for a total of 11520\,s of exposure. The data were collected during gray time, and the seeing varied in the range $0.6$--$0.9^{\prime\prime}$. The data reduction, including bias subtraction and flat-fielding, was performed using standard routines of the Image Reduction and Analysis Facility (IRAF) package, and the cosmic rays were removed with the L.A.Cosmic algorithm \citep{lacos}. We used a set of {\it Gaia} DR3 stars in the target vicinity for the astrometric calibration. The formal RMS uncertainties of the astrometric solution were $\Delta$RA~$\sim$ $0.06^{\prime\prime}$  
and $\Delta$Dec~$\sim$ $0.11^{\prime\prime}$. 
We determined the photometric zero-point $28.43 \pm 0.01$ using several stars from the 
Pan-STARRS catalogue \citep{flewelling_2020_apjs}. To convert their magnitudes to the Sloan Digital Sky Survey (SDSS) photometric system, we used the transformation equations from \citet{tonry_pan}.

The resulting combined image is presented on the right hand side of Figure~\ref{fig:FOV_optical}. At the position of \rtwelve, we detect a faint unresolved source with $r = 27.32 \pm0.16$\,mag. For both the GTC and Binospec images, we perform photometry on sources O1 through O9 in Figure~\ref{fig:FOV_optical}, which are listed in Table~\ref{tab:host}, together with the priors, and results of our PATH analysis.

\begin{table*}
    \caption{Properties of the Host and Surrounding Galaxies of \rtwelve}
    \centering
    \begin{tabular}{lccccc} \hline
        Galaxy & $r$ & Offset ($^{\prime\prime}$) & $P(O)$ &  $P(O\vert x)$ & redshift\\ \hline
        O1 & $23.23\pm0.02$ & 8.97 & 0.054 & 0.0 \\
        O2 & $26.02\pm0.05$ & 8.72 & 0.006 & 0.0 \\
        O3 & $26.72\pm0.09$ & 4.95 & 0.004 & $\approx0$  \\
        O4 & $27.17\pm0.16$ & 0.10 & 0.003 & 0.9995 \\
        O5 & $25.69\pm0.04$ & 2.11 & 0.008 & 0.00017 \\
        O6 & $25.01\pm0.05$ & 3.72 & 0.006 & $\approx0$ \\
        O7 & $24.80\pm0.02$ & 4.27 & 0.015 & $\approx0$ \\
        O8 & $20.31\pm0.02$ & 8.70 & 0.790 & 0.00028 & $0.1935\pm0.0203$\\
        O9 & $23.05\pm0.05$ & 7.74 & 0.063 & 0.0 & $0.5473\pm0.0004$\\ \hline
    \end{tabular} 
\\
    \label{tab:host}
\end{table*}

\subsection{Host Galaxy Association}
\label{sec:host_ass}
We use the Probabilistic Association of Transients to their Hosts algorithm (PATH; \citealt{path}) to determine the most likely host galaxy. PATH is a Bayesian framework that incorporates priors on the magnitudes of surrounding galaxies, their sizes, and the transient's offset from them to calculate posteriors of association for all candidate galaxies, $P(O\vert x)$. We use the GTC/OSIRIS $r$-band image for this analysis, because it is the deepest image available for this field. For the magnitude prior, we use the default `inverse' prior which gives higher weight to brighter galaxies. For the offset prior, we use the default `exponential' prior truncated at six effective radii from the galaxy candidates. Finally, we assume a value of 0.05 for the prior that the host is undetected, $P(U)$, which is a conservative assumption given the depth of the GTC image. 

To select galaxy candidates, we consider a $20^{\prime\prime} \times 20^{\prime\prime}$ region around \rtwelve and use {\tt Source Extractor} \citep{source_extractor} to identify all objects that are likely galaxies per the {\tt Source Extractor} star-galaxy classifier, resulting in nine objects that we label O1--O9. We run PATH using the positions determined by {\tt Source Extractor}, photometry as described in Table~\ref{tab:host}, and estimations for the effective radii of the objects combined with our prior assumptions. O4, the object spatially coincident with the FRB source, is unambiguously favored as the host with $P(O\vert x)$ = 0.9995; all other objects receive negligible posteriors, and the posterior on the host being undetected, $P(U\vert x)$, is similarly negligible at $5\times10^{-5}$. We report the $P(O\vert x)$ values for each object in Table~\ref{tab:host} along with their magnitudes, spectroscopic redshift (if known), and offset from \rtwelve.

The host galaxy of \rtwelve is too faint to obtain a redshift using current ground-based spectroscopy; this might yet still be feasible with the \textit{HST} or \textit{JWST}. To determine a conservative upper limit on the redshift, which would allow us to constrain the maximum possible host galaxy luminosity, we calculate $P$($z\vert$DM) following \citet{macquart_2020_natur}. Throughout we use the model parameter values used in \citet{james_2020_mnras} and Planck18 cosmological parameters \citep{planckcollaboration_2020_aa}. We account for the expected Galactic DM using a flat distribution with a $\pm20$\,\% spread, centred on $71.5$\,pc\,cm$^{-3}$ \citep[from NE2001p;][]{ocker_2024_rnaas}, the Galactic halo contribution is a flat distribution between 25 and 80\,pc\,cm$^{-3}$ \citep{prochaska_2019_mnras,yamasaki_2020_apj}, and we fix the host DM to $0$\,pc\,cm$^{-3}$ in order to place the most conservative limit on the maximum redshift. The field around \rtwelve (Figure~\ref{fig:FOV_optical}) is crowded with other galaxies, implying that there might be significant contribution to the DM from intervening circum-galactic media. We account for this by introducing scatter in the DM component from the inter-galactic medium to make the final probability density function (PDF). The PDF is plotted in Figure~\ref{fig:gallum} (in orange) with the 99\,\% confidence interval defining the maximum redshift: $z_{\mathrm{max}}\sim0.83$. The mean of this distribution, which conservatively assumes no host contribution to the DM, is: $z\sim0.67$. 

Note that scattering measurements can be used to constrain the host DM contribution, under the assumption that the scattering is dominated by the host galaxy disk \citep{cordes_2022_apj}. In our case, we have not measured a scattering timescale in the \rtwelve bursts, but place an upper limit of $<0.65$\,ms using the shortest duration burst, B2. This upper limit is, unfortunately, not meaningfully constraining for the range of host DMs in our case, using a flat prior on the fluctuation parameter between 0.5 and 2 (pc$^2$ km)$^{-1/3}$. We compare the PDF determined following \citet{macquart_2020_natur}, plotted in orange, on Figure~\ref{fig:gallum} with the PDF calculated following the \citet{cordes_2022_apj} framework, plotted in pink, using our upper limit on the scattering timescale. This highlights that the upper limit on the redshift results in a comparable galaxy luminosity ($\sim10^8L_{\odot}$) regardless of the assumed PDF (the \citet{macquart_2020_natur} PDF is skewed low since it accounts for some DM contribution from intervening structure, which the \citet{cordes_2022_apj} framework does not consider). In contrast, the lower bound on the redshift is unconstrained, since the scattering timescale does not meaningfully constrain the maximum DM host.

\section{Discussion}
\label{sec:disc}
\subsection{The Importance of Precise Localization and Deep Optical Observations}

\begin{figure}
    \centering
    \includegraphics[width=0.5\textwidth]{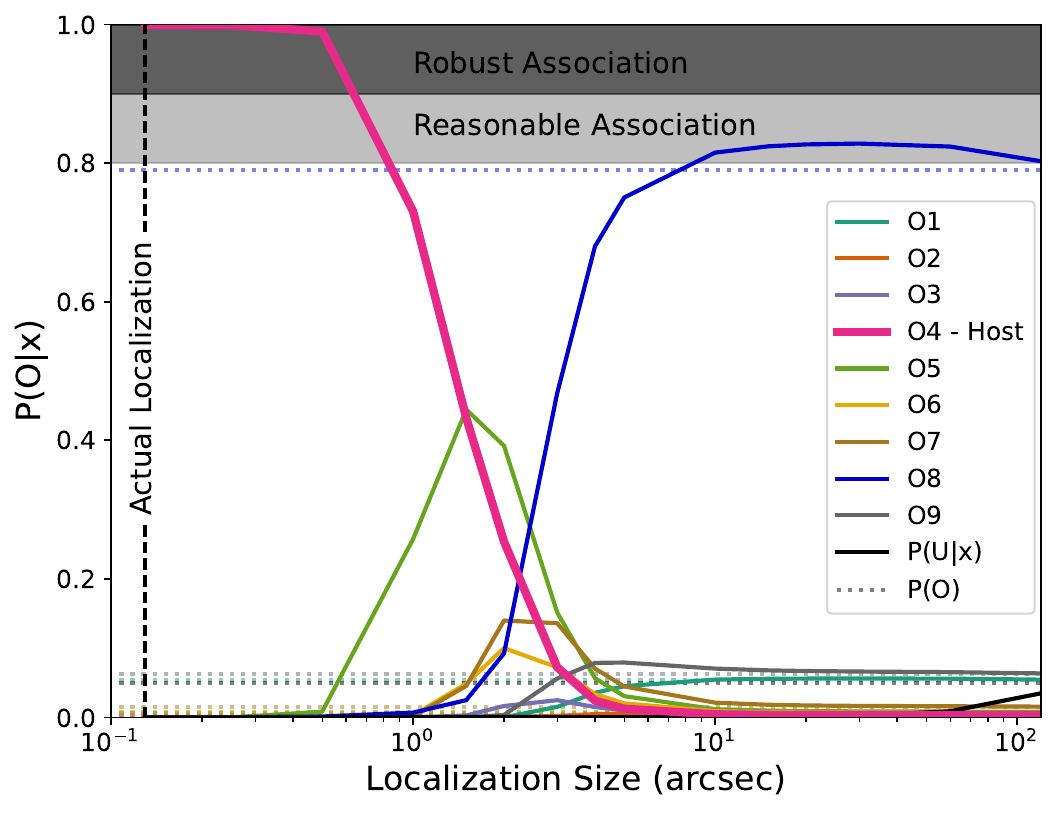}
    \caption{PATH posteriors, $P(O\vert x)$, for the \rtwelve field as a function of increasing assumed localization uncertainty (radius-equivalent). The actual 1-$\sigma_x$ localization precision of the source is shown by the dashed vertical line. The $P(O\vert x)$ for objects O1 to O9 are represented by solid curves and their priors, P(O), by horizontal dotted lines with corresponding colors (see legend). Shaded gray regions illustrate reasonable and robust host associations (defined as $P(O\vert x)>$ 0.8 and $> 0.9$, respectively).}
    \label{fig:path}
\end{figure}

The unambiguous association of \rtwelve to its host is due to both its 260-mas localization and the depth of the GTC image used for the PATH analysis, the combination of which are not always afforded for FRBs. To explore the sensitivity of the posteriors to the localization uncertainty for \rtwelve, we artificially increased the localization uncertainty region from $0.5^{\prime\prime}$ to $120^{\prime\prime}$, while keeping all other assumptions fixed, and noted how the posteriors changed. Figure~\ref{fig:path} shows the $P(O\vert x)$ for each candidate host galaxy (with O4 being the true host) and the $P(U\vert x)$ as a function of localization uncertainty (radius-equivalent).

Figure~\ref{fig:path} shows that for localization sizes $\lesssim 1\arcsec$, O4 is correctly identified as the true host with $P(O\vert x) \gtrsim 0.8$. For a robust association, which we define as $P(O\vert x) \gtrsim 0.9$, a localisation precision on the order of a few 100\,mas, or less, is needed. Once the localization size increases to beyond a few arcseconds, the host association becomes less clear, as evident by several candidates with low but comparable $P(O\vert x)$. At worse precision ($\gtrsim8\arcsec$), the data begin to lose constraining power as $P(O\vert x)$ becomes dominated by the prior; if the localization was truly on this scale, the brightest and largest galaxy, O8, would have been identified as the host with reasonable confidence (defined as $P(O\vert x) \gtrsim 0.8$). This exercise highlights the importance of sub-arcsecond localizations in making robust associations with the true host, especially if a significant fraction of the FRB population originate from faint or low-luminosity galaxies. 

In the case of \rtwelve, the VLBI localization, coupled with the lack of clear host within $\sim 10\arcsec$ in shallower imaging, motivated extremely deep imaging. In previous PATH analyses of this system based on shallower MMT imaging, the FRB appeared significantly offset from all known candidates, as the coincident galaxy was not detected. This demonstrates that, in some cases -- perhaps particularly when the FRB position is significantly offset from its putative host or apparently hostless -- deep optical imaging, in addition to high-precision  localization, may be necessary for a robust host association.

\subsection{The Implications of a Low-Luminosity Dwarf Host}

\begin{figure}
    \centering
    \includegraphics[width=0.5\textwidth]{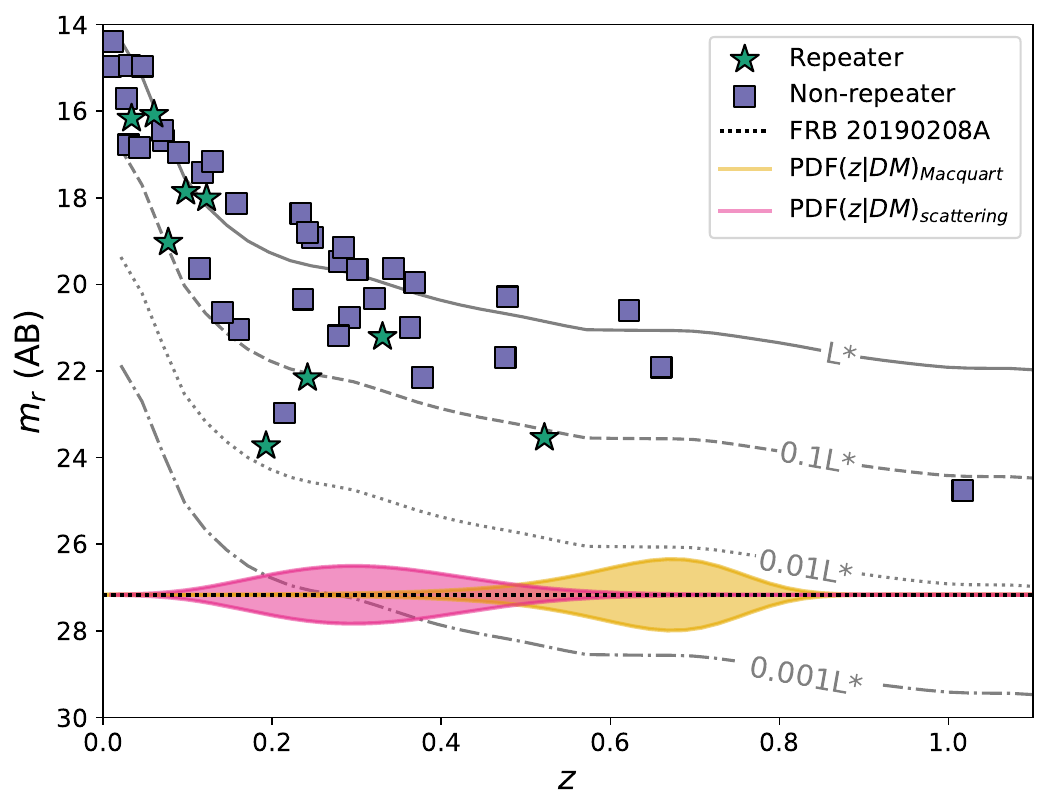}
    \caption{The $r$-band AB magnitude versus inferred redshift for the host of \rtwelve against a background of published FRB host galaxies (references in the main text). We show the placement of L* field galaxies and scale them in orders-of-magnitude down to 0.001L*. The true magnitude of O4 is indicated by the horizontal dashed line. Overplotted on this line are two probability distribution functions (PDFs) for the redshift,  described in detail in Section~\ref{sec:host_ass}. The orange redshift-PDF indicates a conservative estimate, where DM$_{\text{host}}=0\,$pc\,cm$^{-3}$, using the framework from \citet{macquart_2020_natur}. The pink redshift-PDF uses an upper-limit on scattering to constrain DM$_{\text{host}}$, using the framework from \citet{cordes_2022_apj}. The most probable inferred redshift range for \rtwelve, indicates that the host is consistent with a $<$ 0.01L* galaxy.}
    \label{fig:gallum}
\end{figure}

Figure~\ref{fig:gallum} shows the $r$-band AB magnitude against redshift for a sample of published FRB host galaxies, including data from \citealt{gordon_2023_apj,leewaddell_2023_pasa,2024ApJ...971L..51B,ibik_2024_apj,panther_2023_mnras,ravi_2019_natur,bhardwaj_2021_apjl,law_2024_apj,gordon_2024_apjl,rajwade_2024_mnras}. Overplotted is the characteristic luminosity, L*  as a function of redshift, based on the Schechter galaxy luminosity function of field galaxies \citep{schechter_1976_apj}. Also shown are such curves scaled down in orders-of-magnitude, with data compiled from \citealt{Brown+01,Wolf+03,Willmer+06,Reddy+09,Finkelstein+15,heintz_2020_apj}. 

The host galaxy of \rtwelve is very faint, less luminous than the host galaxy of \rone \citep[$\sim0.01$L*,][]{tendulkar_2017_apjl,bassa_2017_apjl}, and potentially as faint as a $\sim0.001$L* galaxy, depending on the true redshift. Considering the extremes, for $z_{\mathrm{O8}}=0.1935$ and $z_{\mathrm{max}}\sim0.83$,  the corresponding luminosities are $\sim10^{6.8}\,L_{\odot}$ and $\sim10^{8.3}\,L_{\odot}$, respectively. The plausible range of redshifts of the host, thus imply that O4 is the lowest-luminosity FRB host galaxy to date. In future, if the redshift of O4 can be determined, it will be possible to comment on whether this host galaxy is an outlier in the FRB population, or merely an extreme case still consistent with the current population of (repeating) FRB host galaxies.

In addition to \rtwelve and \rone, there are a few other low-luminosity FRB hosts (for both repeating and apparently non-repeating sources), such as the hosts of \ronetwin \citep{niu_2022_natur}, and FRB~20210117A \citep{bhandari_2023_apj}. Furthermore, no host galaxy was detected for FRB~20210912A, discovered by ASKAP, despite deep optical ($R>26.7$\,mag) and near-infrared ($K_s>24.9$\,mag) follow-up observations  \citep{marnoch_2023_mnras}. These authors concluded that the host galaxy of FRB~20210912A is either intrinsically dim, or the burst was exceptionally bright. If the source is situated at $z<0.7$, it would be fainter than any FRB host galaxy previously detected. Alternatively, the host could be as, or more luminous than that of \rone, if situated further away at $z>0.7$. The low luminosity of the host of \rtwelve presented in this work, strengthens the feasibility of the scenario where the host of FRB~20210912A is a low-luminosity dwarf galaxy.

Recently it has been shown that there is a significant deficit of low-mass FRB host galaxies in the local Universe \citep{sharma_2024_arxiv}. This implies that high metallicity may play a crucial role in the production of FRB progenitors, and that FRBs may come from magnetars formed in a sub-population of core-collapse supernovae (CCSNe). While various studies have shown that FRBs do not track stellar mass \citep{heintz_2020_apj,bhandari_2022_aj,sharma_2024_arxiv}, the association of \rtwelve with a faint dwarf galaxy, is still consistent with the host population of CCSNe, which have been proposed to be the dominant formation channel for FRB sources in the local Universe \citep{2024ApJ...971L..51B}. For example, \cite{2020ApJ...904...35P} found that approximately 7\,\% of CCSNe in the Zwicky Transient Facility (ZTF) Bright Transient Survey occur in very low-luminosity galaxies (absolute i-band magnitude M$_i > -16$\,mag). Notably, the faintest Type~II host galaxy in their sample, SN2024rnu at $z = 0.032$, has M$_i = -11.84$\,mag, comparable to the lower magnitude limit estimated for the host of \rtwelve. Similar findings have been reported by \cite{2010ApJ...721..777A} and \cite{2021MNRAS.503.3931T} using unbiased samples of nearby CCSN host galaxies from the Palomar Transient Factory (PTF) and the All-Sky Automated Survey for Supernovae (ASAS-SN). 
 
Alternatively, this FRB source may be one of a few outlier cases, associated with extreme transients like superluminous supernovae (SLSNe) and long gamma-ray bursts (LGRBs), which predominantly occur in dwarf galaxies with high specific star-formation rates \citep[e.g.,][]{fruchter_2006_natur,savaglio_2009_apj,schulze_2018_mnras}. The idea that FRBs may be linked to extremely massive progenitor stars, thought to be responsible for these extreme transients, was already put forward when the host galaxy of \rone was identified as a low-metallicity dwarf \citep{tendulkar_2017_apjl}. The magnetars, hypothesized to power H-poor SLSNe \citep{kasen_2010_apj}, could potentially produce FRBs as well \citep{metzger_2017_apj}. The preference these extreme transients have for low-mass galaxies signals that these galaxies provide certain conditions \citep[e.g., in terms of metallicity;][]{schulze_2018_mnras} that are conducive to their production. At present, we lack sufficient evidence to strongly favour either scenario.

Finally, it is worth noting that the environments of the least massive, star-forming galaxies are representative of the earliest starburst galaxies in the Universe. The discovery of multiple FRBs in such galaxies could thus bode well for high-redshift FRB searches.

\subsection{The Lack of a Compact Persistent Radio Source}

Two sources of FRBs are coincident with compact persistent radio sources (PRSs) with flat spectra and spectral luminosity $\sim10^{29}\,$erg\,s$^{-1}$\,Hz$^{-1}$: \rone\ \citep{marcote_2017_apjl} and \ronetwin\ \citep{niu_2022_natur,bhandari_2023_apjl}. These FRBs also have remarkably high RM values. The PRSs have been hypothesized to be magnetized neutron stars embedded in SN remnants, or wind nebulae \citep[e.g.,][]{margalit_2018_apjl}, or ultraluminous X-ray source hypernebulae \citep{sridhar_2022_apj}. Notably, some of the most active repeaters to date (\rsixtyseven, \rnarwhal, FRB~20240114A), do not exhibit a PRS of similar luminosity \citep{nimmo_2022_apjl,hewitt_2024_mnras,kumar_2024_arxiv_arxiv240612804}, although \citet{bruni_2023_arxiv} have detected a possible low-luminosity ($\sim10^{27}\,$erg\,s$^{-1}$\,Hz$^{-1}$) PRS associated with \rsixtyseven.  

This low luminosity is consistent with the theoretical relation predicted between the luminosity of a PRS and RM, if the PRS is the main contributor to the RM \citep{yang_2020_apj}. However, this low-luminosity PRS is quite different from those associated with \rone or \ronetwin. Constraints on the size are a hundred times larger, the spectrum is inverted and not flat, and the host galaxy is a barred-spiral as opposed to a dwarf galaxy \citep{xu_2022_nature,dong_2024_apj}. 

We performed a search for a PRS associated with \rtwelve using our EVN observations at 1.382\,GHz, as well as the VLA in C-configuration at 6\,GHz. With the EVN, we found no persistent radio emission on milliarcsecond scales in the $2\times2$\,arcsec$^2$ region surrounding the position of \rtwelve.  The image had an RMS of 31\,$\upmu$Jy\,beam$^{-1}$, resulting in a 5-$\sigma$ upper limit of 155$\upmu$Jy\,beam$^{-1}$ (see the left panel of Figure~\ref{fig:contmaps}). Similarly using the VLA image (4\,$\upmu$Jy\,beam$^{-1}$ RMS), we could rule out the presence of a PRS above a 3-$\sigma$ flux density limit of 12\,$\upmu$Jy\,beam$^{-1}$ (right panel of Figure~\ref{fig:contmaps}). A PRS of spectral luminosity $2\times10^{29}$\,erg\,s$^{-1}$\,Hz$^{-1}$ (comparable to those of \rone and \ronetwin), would have a flux density of $\sim420\,\upmu$Jy at $z_{\mathrm{O8}}=0.1935$,and $\sim10\,\upmu$Jy at $z_{\mathrm{max}}\sim0.83$. Our data thus rules out the presence of such a luminous PRS associated with \rtwelve assuming reasonable redshifts. 

Given the modest RM of \rtwelve \citep[][Appendix~\ref{sec:pol_eff} and \ref{sec:pol_nrt}]{mckinven_2023_apj}, the aforementioned RM-luminosity relation suggests a PRS luminosity $<10^{26}\,$erg\,s$^{-1}$\,Hz$^{-1}$, resulting in an expected flux density well below our detection capability even at the closest feasible redshift.

\subsection{Exploring Alternative Host Scenarios}

At least one repeater, FRB~20200120E \citep{bhardwaj_2021_apjl}, is located in a globular cluster \citep{kirsten_2022_natur}. Globular clusters tend to be within a few tens of kiloparsecs from the center of their host galaxies but can also be over a hundred kiloparsecs away, since they trace the dark matter halo of the host \citep[e.g.,][]{reinacampos_2022_mnras}. Even in the Milky Way, there are globular clusters beyond 100\,kpc from the Galactic centre \citep{harris_1996_aj}. Alternatively, satellite galaxies in the Local Group are typically within a few hundred kiloparsecs of their hosts \citep[see e.g.,][]{mcconnachie_2012_aj}. It thus remains plausible that the host galaxy of \rtwelve is a low-luminosity satellite of another galaxy in the field, or even a (very luminous) globular cluster. We discuss these options only for the galaxies for which we have measured spectroscopic redshifts, O8 and O9, although depending on the redshift of other galaxies in the field, the discussion may be applicable to them as well. 

If \rtwelve is associated with galaxy O8 ($z=0.1935$), the impact parameter is $\sim28$\,kpc. Assuming similar offsets as seen for the Milky Way, this is broadly consistent with the offsets of both satellite galaxies and globular clusters discussed above. At this redshift, a globular cluster would have to be extremely luminous, M$_r=-12.6$\,mag. Perhaps the biggest challenge to association with O8 is the consequent unexplained DM-excess. Having accounted for contributions from the halo and disk of the Milky Way, there would be an excess of  ($\sim300$\,pc\,cm$^{-3}$), comparable to the local DM contributions seen in \rone and \ronetwin. However, the absence of a PRS, and the lack of substantial scattering or Faraday rotation in the bursts from \rtwelve hints at the absence of a significant local environment contribution to the DM. 

In the case of association with O9 ($z=0.547$), the large DM-excess discrepancy would be solved, and the larger impact parameter ($\sim50$\,kpc) would still be consistent with the \rtwelve host being a satellite dwarf galaxy of O9. However, a globular cluster origin becomes implausible given the relatively large offset, and more importantly, the implied luminosity at such a distance (M$_r=-15.2$).

Lastly further opposing a globular cluster origin, although the bursts from FRB~20200120E are similar in terms of narrowbandedness and polarimetry, they are $\sim30$ times shorter in duration and $\sim10^2$ times less luminous than those of other extragalactic FRBs \citep{nimmo_2023_mnras}. It is unknown whether these properties are linked to its globular cluster origin. The \rtwelve bursts, on the other hand, have more typical durations and spectral energies \citep[see also][]{fonseca_2020_apjl}.

\section{Summary}
Using the EVN in EVN-Lite mode, we monitored \rtwelve over three years, accumulating 65.6\,hrs of observations as part of the PRECISE campaign, and detected a single burst (B1) during this period. With the \'ECLAT monitoring campaign on the NRT, we obtained an additional 40.4\,hours of exposure at $\sim1.4$\,GHz, leading to the detection of one more burst (B2).

The detection of B1 enabled the VLBI localization of \rtwelve to RA (J2000) = $18^{\rm h}54^{\rm m}11\fs27\pm 260\,$mas Dec (J2000) = $+46^{\circ}55^{\prime}21\farcs67\pm 260\,$mas. Initial optical observations with the MMT ($3\sigma$ limit of $i \gtrsim 25.7$\,mag (AB)) revealed no host at the FRB position. Follow-up observations with the GTC revealed a faint source ($r = 27.32 \pm0.16$\,mag) at the position of \rtwelve. Even the most conservative redshift estimate inferred from the DM,  $z_{\mathrm{max}}\sim0.83$, indicates that the \rtwelve host is less luminous that the dwarf host galaxy of \rone, making it the lowest-luminosity FRB host galaxy to date. 

There has been an emerging trend where many FRB host galaxies are massive and star-forming. A low-luminosity FRB host galaxy, such as presented here, might still be consistent with the framework where the majority of FRBs are produced by magnetars formed via CCSNe, but could also be part of a sub-population where the FRB progenitor is linked to extreme transient events such as LGRBs or SLSNe (requiring low metallicity).

In the coming decade, the number of host associations will drastically increase with the advent of large-scale localization projects such as the CRACO upgrade on the Australian Square Kilometre Array Pathfinder (ASKAP) \citep{shannon_2024_arxiv}, the Deep Synoptic Array \citep[DSA; ][]{hallinan_2019_baas}, the Canadian Hydrogen Observatory and Radio-transient Detector \citep[CHORD; ][]{vanderlinde_2019_clrp}, and CHIME/FRB outriggers \citep{lanman_2024_aj}. Many scenarios may arise where localization regions are offset from potential host galaxies. In some cases, robust FRB host associations will require both high precision localisation and deep optical follow-up observations.

\vspace{2cm}
\centerline{ACKNOWLEDGEMENTS}

We thank Betsey Adams (ASTRON) for useful discussions.

The AstroFlash research group at McGill University, University of Amsterdam, ASTRON, and JIVE is supported by: a Canada Excellence Research Chair in Transient Astrophysics (CERC-2022-00009); the European Research Council (ERC) under the European Union’s Horizon 2020 research and innovation programme (`EuroFlash'; Grant agreement No. 101098079); and an NWO-Vici grant (`AstroFlash'; VI.C.192.045).

Y.D., W.F., A.C.G., and the Fong Group at Northwestern acknowledge support by the National Science Foundation under grant Nos. AST-1909358, AST-2308182 and CAREER grant No. AST-2047919. Y.D., W.F., A.C.G., and L.M.R. acknowledge support from NSF grants AST-1911140, AST-1910471 and AST-2206490 as members of the Fast and Fortunate for FRB Follow-up team. A.K. acknowledges the DGAPA-PAPIIT grant IA105024. K.N. is an MIT Kavli fellow. 

S.B. is supported by a Dutch Research Council (NWO) Veni Fellowship (VI.Veni.212.058). W.F. gratefully acknowledges support by the David and Lucile Packard Foundation, the Alfred P. Sloan Foundation, and the Research Corporation for Science Advancement through Cottrell Scholar Award \#28284. A.K. acknowledges the DGAPA-PAPIIT grant IA10502. A.GdP. acknowledges financial support from the Spanish Ministerio de Ciencia e Innovación under grant PID2022-138621NB-I00. F.K. acknowledges support from Onsala Space Observatory for  the  provisioning of its facilities/observational support. The Onsala Space Observatory national research infrastrcuture is funded through Swedish Research Council grant No 2017-00648. B.M. acknowledges financial support from the State Agency for Research of the Spanish Ministry of Science and Innovation, and FEDER, UE, under grant PID2022-136828NB-C41/MICIU/AEI/10.13039/501100011033, and through the Unit of Excellence Mar\'ia de Maeztu 2020--2023 award to the Institute of Cosmos Sciences (CEX2019-000918-M).

Y.D. is supported by the National Science Foundation Graduate Research Fellowship under Grant No. DGE-1842165. V.M.K. holds the Lorne Trottier Chair in Astrophysics \& Cosmology, a Distinguished James McGill Professorship, and receives support from an NSERC Discovery grant (RGPIN 228738-13), from an R. Howard Webster Foundation Fellowship from CIFAR. K.W.M. holds the Adam J. Burgasser Chair in Astrophysics and is supported by NSF grants (2008031, 2018490). A.B.P. is a Banting Fellow, a McGill Space Institute~(MSI) Fellow, and a Fonds de Recherche du Qu\'ebec -- Nature et Technologies~(FRQNT) postdoctoral fellow. K.R.S. and V.S are supported by FRQNT Doctoral Research Awards. K.S. is supported by the NSF Graduate Research Fellowship Program. This work was supported by the National Natural Science Foundation of China (Grant Nos. 12041304 12288102)

This work is based in part on observations carried out using the 32-m radio telescope operated by the Institute of Astronomy of the Nicolaus Copernicus University in Toru\'n (Poland) and supported by a Polish Ministry of Science and Higher Education SpUB grant. This work is also based in part on observations carried out using the 32-m Badary, Svetloe, and Zelenchukskaya radio telescopes operated by the Scientific Equipment Sharing Center of the Quasar VLBI Network (Russia). We thank the directors and staff at the various participating antenna stations for allowing us to use their facilities and running the observations. The European VLBI Network (EVN) is a joint facility of independent European, African, Asian, and North American radio astronomy institutes. Scientific results from data presented in this publication are derived from EVN project code EK050.

The Nan\c{c}ay Radio Observatory is operated by the Paris Observatory, associated with the French Centre National de la Recherche Scientifique (CNRS). 

Observations reported here were also obtained at the MMT Observatory, a joint facility of the University of Arizona and the Smithsonian Institution. MMT Observatory access was supported by Northwestern University and the Center for Interdisciplinary Exploration and Research in Astrophysics (CIERA). Part of this work is based on observations made with the Gran Telescopio Canarias (GTC), installed at the Spanish Observatorio del Roque de los Muchachos of the Instituto de Astrof\'{i}sica de Canarias, on the island of La Palma. 

Lastly, part of this research has made use of the EPN Database\footnote{\url{http://www.jodrellbank.manchester.ac.uk/research/pulsar/Resources/epn/}} of Pulsar Profiles maintained by the University of Manchester.


%

\vspace{5mm}
\facilities{EVN, NRT, VLA, MMT, GTC}


\software{AIPS \citep{greisen_2003_assl}, astropy \citep{AstropyCollaboration_2013_AA}, CASA \citep {mcmullin_2007_aspc,vanbemmel_2022_pasp}, difmap\citep{shepherd_1994_baas}, DM\_phase \citep{seymour_2019_ascl}, DSPSR \citep{vanstraten_2011_pasa}, FETCH \citep{agarwal_2020_ascl}, heimdall \footnote{\url{sourceforge.net/projects/heimdall-astro/}}, IonFR \citep{sotomayorbeltran_2013_ascl}, PATH \citep{aggarwal_2021_apj}, PRESTO \citep{ransom_2001_phdt}, PolConvert \citep{martividal_2016_aa}, POTPyRI\footnote{\url{https://github.com/CIERA-Transients/POTPyRI}}, PSRCHIVE \citep{hotan_2004_pasa}, PypeIt \citep{pypeit:zenodo,pypeit:joss_pub}, scipy \citep{Virtanen_2020_NatMe}, SFXC \citep{keimpema_2015_exa}, Source Extractor \citep{source_extractor}}



\appendix

\section{EVN observations and data processing}

\begin{figure}
    \centering
    \includegraphics[width=0.5\textwidth]{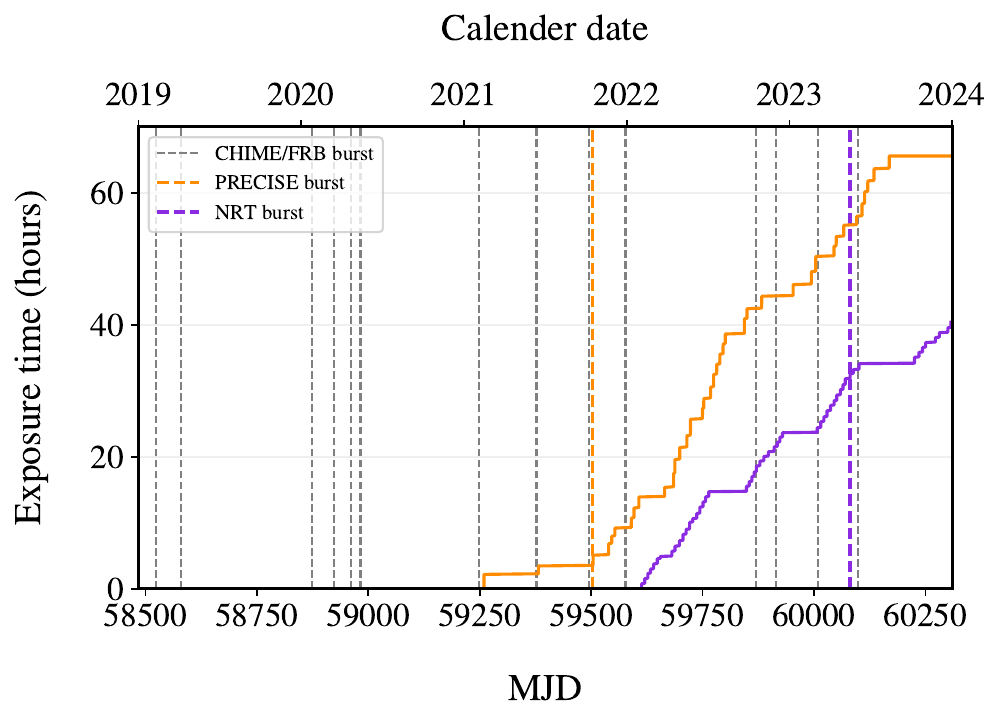}
    \caption{All known bursts detected from \rtwelve are indicated by dashed vertical lines. Bursts detected by CHIME/FRB are indicated in gray, while the bursts detected by PRECISE (B1) and \'ECLAT (B2) are in orange and purple, respectively. The MJD is shown on the bottom axis and the corresponding calendar date on the top axis. The solid orange curve shows the cumulative time (in hours) that \rtwelve has been observed by PRECISE, while the solid purple shows the same but for the \'ECLAT project on the NRT.}
    \label{fig:R12_timeline}
\end{figure}

\label{app:evn}

We have provided an observation log as a supplementary text file which shows all the \rtwelve observations conducted with a sub-array of EVN dishes in `EVN-Lite' mode as part of the PRECISE FRB localization campaign. During all 38 observations only a single burst (B1) was detected on 2021 October 17~UT (MJD~59504) in PRECISE observing run PR187A. This is also shown in Figure~\ref{fig:R12_timeline}.

\begin{figure}
    \centering
    \includegraphics[width=0.48\textwidth]{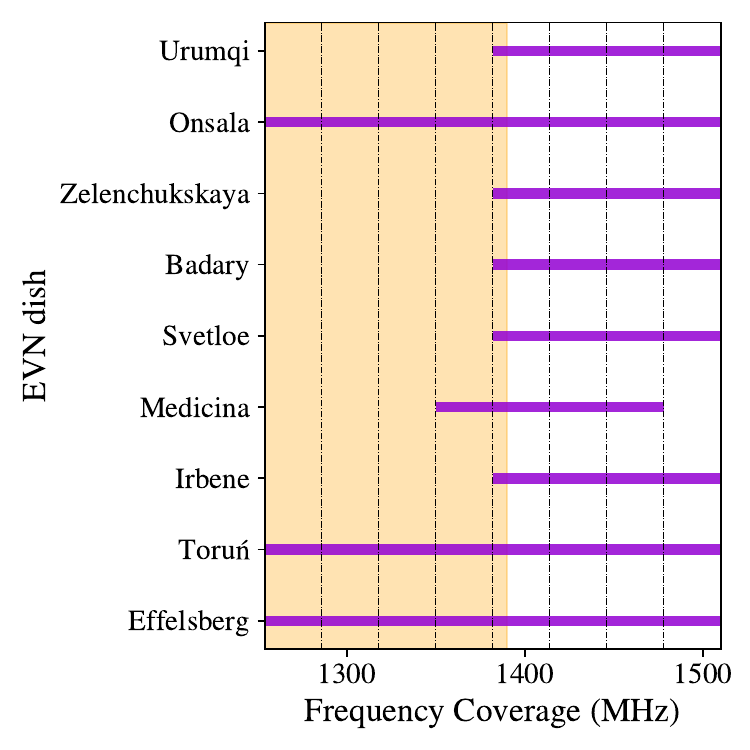}
    \caption{The frequency coverage for different antennas in our PRECISE observations are shown here by horizontal purple bars. The edges of sub-bands are indicated by dashed black vertical lines. The shaded orange region indicates the spectral extent of burst B1.}
    \label{fig:freq_coverage}
\end{figure}

During PR187A, most stations recorded dual-polarization raw-voltage data with 2-bit sampling. At Irbene only left circular polarization data were recorded. These data were in MARK5B \citep[][]{whitney_2004_evn} format for Svetloe, Badary and Zelenchukskaya, and VDIF \citep[VLBI Data Interchange Format;][]{whitney_2010_ivs} for all other stations. With the exception of Urumqi, all stations record in a circular basis. In post-processing, linear basis polarization data from Urumqi were transformed to circular basis using the \texttt{PolConvert} program \citep{martividal_2016_aa}. All stations recorded either four or eight 32-MHz subbands, as shown in  Figure~\ref{fig:freq_coverage}. Unfortunately, B1 occurred in part of the band with less than optimal coverage, as shown by the shaded orange region of this figure.

We used the PRECISE pipeline \citep{kirsten_2021_natas} to search the raw voltages recorded at Effelsberg for bursts. Firstly, we used \texttt{digifil} from the software suite DSPSR \citep[Digital Signal Processing Software for Pulsar Astronomy;][]{vanstraten_2011_pasa} to create Stokes~I filterbank data from the raw-voltage data, with time and frequency resolutions of 64\,$\upmu$s and 62.5\,kHz, respectively. We then searched these filterbank data, over a DM range of $529-629$\,pc\,cm$^{-3}$, for transient signals above a signal-to-noise ratio (S/N) threshold of 7 using the transient-detection software \texttt{Heimdall}\footnote{\url{sourceforge.net/projects/heimdall-astro/}}. Thereafter, the pipeline fed the candidates from the search to \texttt{FETCH} \citep{agarwal_2020_ascl} -- a machine learning convolutional neural network that assigns a probability that a given transient signal is astrophysical in origin, using various deep-learning models. In the PRECISE project, we have empirically determined that the \texttt{FETCH} models `A' and `H' complement each other well in terms of completeness and number of false positives. Finally, we manually inspected all the candidates for which either of these two models assigned a probability of higher than 0.5 of being astrophysical.

In the first correlation pass at JIVE with \texttt{SFXC} \citep{keimpema_2015_exa}, all non-target scans were correlated with the standard 2-s integration time and 8\,$\times$32-MHz sub-bands, consisting of 64 channels each. These data were used to provide an accurate calibration that could be applied to the burst data correlated in the following passes. Initially, \rtwelve's position (derived from the CHIME/FRB baseband data) was only known with arc-minute precision, and we used this position as the phase center in our initial correlations: RA (J2000) = $18^{\rm h}54^{\rm m}09\fs4320$, Dec (J2000) = $46^{\circ}55^{\prime}34\farcs680$. 
In order to first establish the burst position to an uncertainty of about $1^{\prime\prime}$, burst B1 was coherently dedispersed (intra-channel dedispersion) using a DM of 580.467\,pc\,cm$^{-3}$, before a correlation gate was selected manually around the arrival time of B1 to optimise the S/N.
We then derived the delay residuals for each baseline by fringe-fitting these data; the delay residuals are proportional to the angular offset between the burst position and the phase center. This method is known as delay-mapping. A detailed description of this technique has previously been presented in \citet{marcote_2020_natur}. Using the delay-mapping position as the phase center, a second correlation pass was conducted on the burst data, to finally image the burst.  In the third and final correlation pass, all target scans were correlated using the same time and frequency integration as for the calibrators for deep imaging, to search for persistent radio emission, using the same position for the target as derived from the burst correlation.

To calibrate and image the correlated EVN data, we used \texttt{AIPS} \citep{greisen_2003_assl} and \texttt{DIFMAP} \citep{shepherd_1994_baas}. These procedures have also been described in previous PRECISE localization papers \citep[e.g.,][]{marcote_2020_natur}, but we repeat them here for the sake of completeness and convenience. The EVN provides the correlated visibilities in FITS-IDI format, together with calibration pipeline products. We first loaded these correlated visibilities into \texttt{AIPS} before applying the calibration table containing the {\it a-priori} gain correction and parallactic angle correction, the {\it a-priori} flagging table and the bandpass calibration table. We then manually flagged the RFI in the fringe finder (J1419+5423) scan and the edges of each sub-band where the signal becomes weaker ($\sim\,15$\,\% of the data). Next, we corrected for ionospheric dispersive delays using Jet Propulsion Laboratory maps of total electron content during the observation at the different EVN sites. Each sub-band has a different signal path which induces phase jumps between sub-bands, as well as phase slopes within them. With Effelsberg as a reference antenna, we used the fringe finder scan to correct for these instrumental delays. Thereafter, we performed a global fringe fit, using the fringe finder and phase calibrator (J1852+4855), to correct the delays and rates of the phases, as a function of both time and frequency, for all the calibrators during the entire observation. We manually inspected the solutions and flagged those where the calibration failed. These solutions were applied to the check source (J1850+4959) for verification, and then to the target field before imaging. 

In previous PRECISE localizations, the uncertainty on the FRB position was taken to be the quadrature sum of multiple factors: the statistical uncertainty derived from the shape and size of the synthesized beam normalized by the S/N; the statistical uncertainty on the position of the phase calibrator; an estimate of the uncertainty from phase-referencing due to the angular separation between the phase calibrator and FRB; an estimate of the frequency-dependent shift in the phase calibrator position from the International Celestial Reference Frame (ICRF); and the statistical uncertainty on the positions of the interferometric check source. All of these contributions are typically on the order of milliarcseconds or less, and the total uncertainty was often dominated by the phase-referencing uncertainty estimate. 

This approach is effective when the visibilities of multiple bursts can be stacked to obtain sufficient uv-coverage for a single point source to become visible in the dirty map. The precision to which the FRB position is then known is approximately equivalent to 20\,\% of the synthesized beam size. With poor uv-coverage sidelobes become large in amplitude, and while the addition of amplitudes will cause the S/N of different sidelobes to vary, the addition of phases can cause the entire fringe pattern to shift. This means that the true FRB position is not necessarily on top of a local maximum of a sidelobe. Due to the ambiguity in the case of \rtwelve presented here, we conservatively transform 2$\times$$\sigma_{\rm y}$ of the 2D-Gaussian fit into the RA and Dec reference frame to obtain a quasi-positional error on the FRB position.

\begin{figure*}
    \centering
    \includegraphics[width=.475\textwidth]{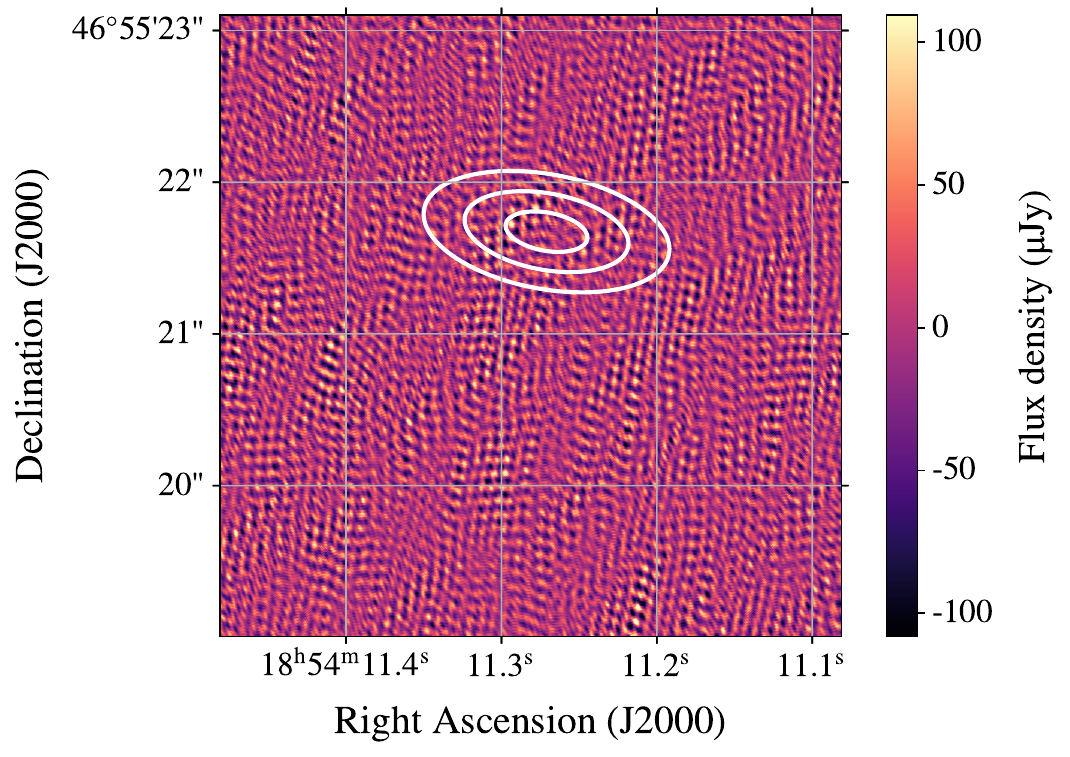}
    \includegraphics[width=.475\textwidth]{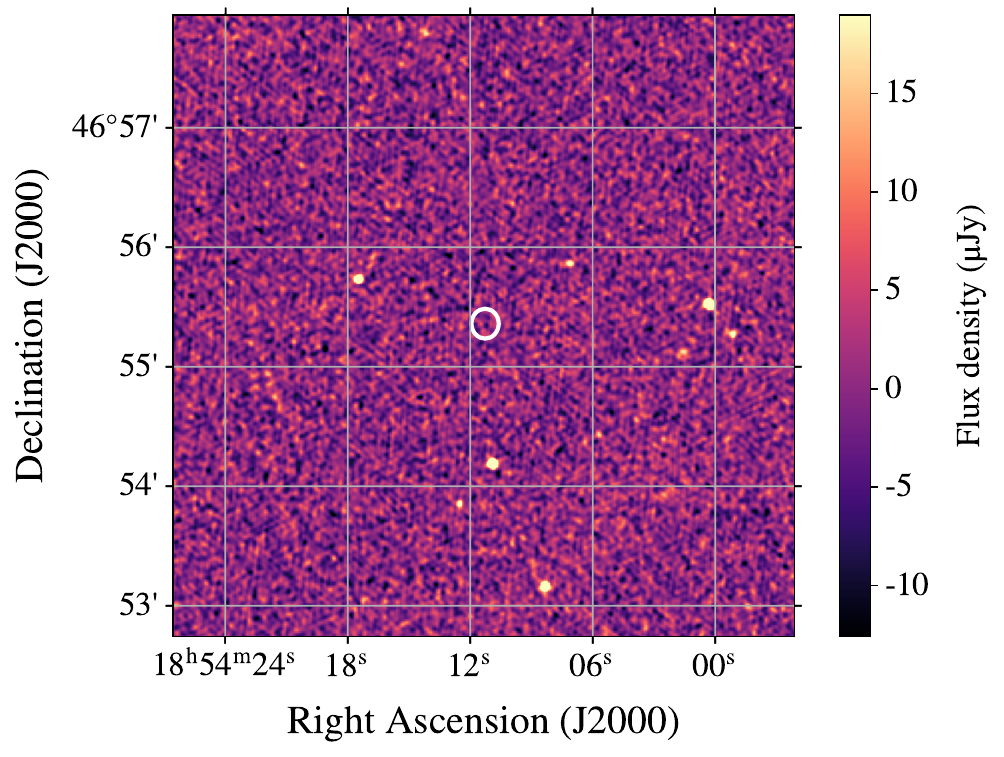}
    \caption{Radio continuum maps from our 1.4-GHz EVN observations (uncleaned, {\bf Left}) and the VLA observations in C-configuration at 6\,GHz (cleaned, {\bf Right}) to search for persistent radio emission. The RMS is 31 and 4\,$\upmu$Jy\,beam$^{-1}$ for the EVN and VLA images, respectively. The ellipses in the EVN image are centered on the VLBI position of \rtwelve and are described in Section~\ref{sec:loc}. The white circle in the VLA image also indicates the FRB position, but is much larger than the aforementioned ellipses. We find no persistent radio sources associated with \rtwelve.}
    \label{fig:contmaps}
\end{figure*}

\section{NRT observations and data processing}
\label{app:nrt}

In the supplementary materials we have provided an observation log of the observations from the \'ECLAT FRB monitoring campaign on the NRT of \rtwelve. Only a single burst (B2) was detected on 2023 May 17~UT (MJD~60081).

The data recorded by NUPPI were searched for bursts using the \'ECLAT pipeline, which has been described in more detail in \cite{hewitt_2023_mnras}. In short, the eight 32-bit, 64-MHz full-polarization subbands recorded by NUPPI were spliced together and converted into 8-bit, 512-MHz Stokes~I filterbank data, which were then passed in 2-minute blocks to \texttt{rfifind} from the pulsar software suite PRESTO \citep{ransom_2001_phdt}. Once the most RFI-contaminated channels were identified, we masked these channels but refrain from any temporal masking. Using \texttt{Heimdall} these masked filterbank data were searched for candidates above a S/N of 7. The DM search range used for \rtwelve is $558-609$\,pc\,cm$^{-3}$ (the data were coherently dedispersed to a DM of 579\,pc\,cm$^{-3}$ within each 4-MHz channel). Notably, we searched for candidates down to the native time resolution of the data (16\,$\upmu$s). Afterwards, the \texttt{FETCH} models `A' to `H' classified the candidates as being astrophysical in origin, and all those candidates for which any model assigned a score above 0.5 were manually inspected. In the event that bursts are detected, they are extracted with full polarization information from the 32-bit raw data. These extracted data are used for all subsequent analyses.

\section{Burst analysis}
\label{app:burpop}

In Table~\ref{tab:properties} we tabulated the properties of bursts B1 and B2. The temporal burst widths and spectral extents quoted are the full-width-at-half-maximum (FWHM) of Gaussian fits to the frequency- and time-averaged data, respectively. We calculate fluences by first multiplying the normalised frequency-averaged burst profiles (where noise has a mean of zero and standard deviation of one) with the radiometer equation to convert to physical units, and then integrating over the extent of the burst in time and the \textit{entire} observing band. The system temperature and gain for the NRT at 1.4\,GHz are approximately 35\,K and 1.4\,K\,Jy$^{-1}$, respectively. The system temperature and gain for Effelsberg are 20\,K and 1.4\,K\,Jy$^{-1}$, respectively. We assume these values, and by extension our fluences, to have an uncertainty of approximately 20\,\%.

\subsection{DM Determination}
\label{sec:dm}

Accurate DM determination in the absence of prominent burst structure or short-time-scale features is non-trivial. In these cases, the DM will likely be overestimated since the intra-burst time-frequency drift \citep{hessels_2019_apjl} cannot easily be disentangled from dispersion, and S/N will be optimised for rather than structure. To determine the DM we made use of both S/N-optimisation and a structure-optimisation approaches. For B1, we have baseband data, and thus generated a filterbank data set, coherently dedispersed to a preliminary DM of 579.8\,pc\,cm$^{-3}$, with 8\,$\upmu$s time resolution and $250$\,kHz frequency resolution. This preliminary DM is similar to the DM of \rtwelve bursts found by CHIME/FRB around the same period of time \citep{mckinven_2023_apj}. For B2 the burst data we used for this analysis had the native time and frequency resolution (16\,$\upmu$s and 4\,MHz) that was recorded at NRT, coherently dedispersed to a DM of 579\,pc\,cm$^{-3}$ used in the search pipeline. For both B1 and B2 we first flag any channels contaminated by RFI before performing the analysis.

\begin{figure*}
    \centering
    \includegraphics[width=.475\textwidth]{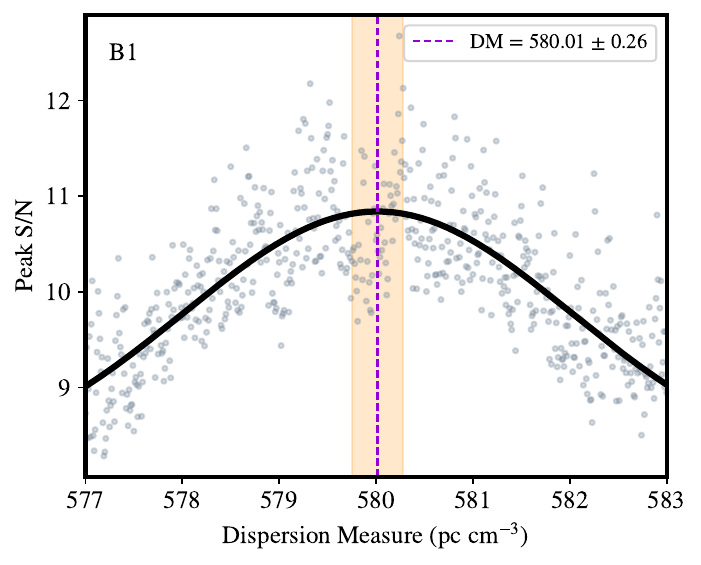}
    \includegraphics[width=.475\textwidth]{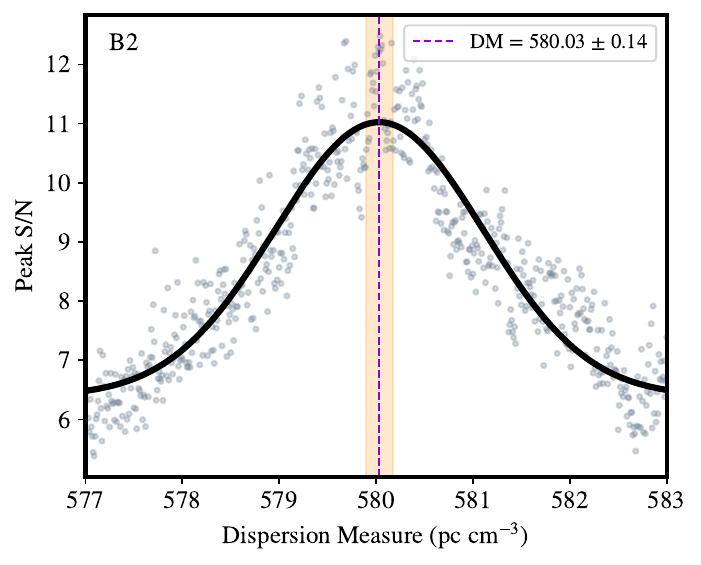}
    \caption{DM determination by S/N optimisation for bursts B1 (\textbf{Left}) and B2 (\textbf{Right}). The gray data points denote the peak S/N at various DM trials. A Gaussian fit to the data is shown in black. The center of the Gaussian and its $3\sigma$ error are indicated by the vertical dashed purple lines and the shaded orange regions, respectively.}
    \label{fig:dm_opt}
\end{figure*}

In order to optimise for S/N, we incoherently dedispersed the bursts to a range of DMs between 577 and 583\,pc\,cm$^{-3}$. For each DM trial, we measured the peak S/N of the profile obtained by averaging the channels where the burst is present. We then performed a bootstrap-fit of a Gaussian function to the S/N vs DM curve, to quantify the best DM and estimate the uncertainty, finding best DMs of $580.01\pm0.26$ and $580.03\pm0.14$ pc\,cm$^{-3}$ for B1 and B2, respectively (Figure~\ref{fig:dm_opt}).

In order to optimise for burst structure, we made use of \texttt{DM\_phase} \citep{seymour_2019_ascl}, and considered only the part of the observing band where the burst is visible while also limiting fluctuation frequencies to below 0.75\,ms$^{-1}$. The resulting DMs for B1 and B2 are $580.24\pm0.23$ and $579.84\pm0.29$\,pc\,cm$^{-3}$, respectively. 

The results from both methods are consistent within the uncertainties, for both bursts B1 and B2, suggesting that both bursts lack any prominent substructure on sub-millisecond time-scales. For the remainder of analyses in this work we will use DMs of 580.1 and 579.9\,pc\,cm$^{-3}$ for bursts B1 and B2, respectively. For the Effelsberg data, the smearing in the lowest 250-kHz channel, if no coherent dedispersion is applied, is 0.6\,ms. We thus create a filterbank data set for B1, coherently dedispersed to a DM of 580.1\,pc\,cm$^{-3}$ to mitigate the DM-smearing. The NRT data for \rtwelve were already coherently dedispersed to a DM of 579\,pc\,cm$^{-3}$ at the time of recording. This difference of 0.9\,pc\,cm$^{-3}$ results in a temporal smearing of $\sim16\,\upmu$s at the lowest frequency channel, which is equivalent to the time resolution of our data, and consequently negligible. 

\subsection{Scattering, Scintillation and High-Time-Resolution Analysis}

FRBs can be asymmetrically temporally broadened through scattering, induced by screens of turbulent media along the line-of-sight to the source \citep[e.g.,][]{nimmo_2024_arxiv}. The estimated extragalactic pulse broadening ($\tau_X$) along the line-of-sight towards \rtwelve is 0.499\,$\upmu$s at 1\,GHz \citep[NE2001p;][]{cordes_2002_arxiv,ocker_2024_rnaas}. Assuming a $\tau\propto\nu^{-4}$ scaling, the corresponding expected scattering timescales at the center frequencies of our PRECISE (1382\,MHz) and NRT (1484\,MHz)  observations are 0.137 and 0.103\,$\upmu$s, respectively. This is much smaller than the time resolution of our NRT observations, and we see no prominent scattering tail in neither B1 nor B2. CHIME/FRB have reported scattering times $<1.8$\,ms at 600\,MHz for \rtwelve bursts \citep{fonseca_2020_apjl}. The scaled expectation from NE2001 is 3.850\,$\upmu$s. 

\begin{figure}
    \centering
    \includegraphics[width=0.5\textwidth]{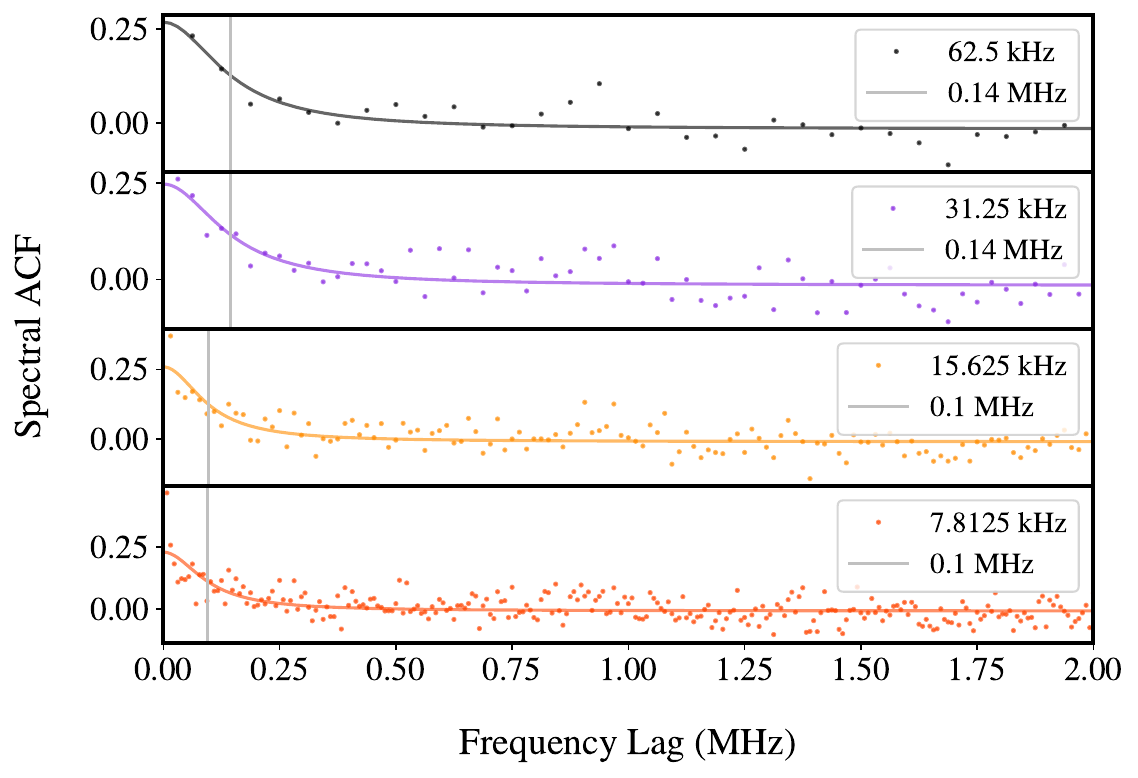}
    \caption{ The time-averaged spectral ACF of B1 at different frequency resolutions (quoted in each panel). Lorentzian fits have been over-plotted in colored lines, while the vertical gray line in each panel indicates the measured scintillation bandwidth for that frequency resolution.  }
    \label{fig:R12_freqacf}
\end{figure}

Alternatively, these scattering screens can also induce scintillation in the FRB spectra. While there is also no prominent scintillation visible by eye in the dynamic spectra of B1 and B2, the expected scintillation bandwidth at 1\,GHz along this line-of-sight is 0.37\,MHz \citep[NE2001p;][]{cordes_2002_arxiv,ocker_2024_rnaas}. Scaling (using $\nu_{\rm{sb}}\propto\nu^{4}$) to the central frequencies of our observations yields 1.35\,MHz at 1382\,MHz and 1.79\,MHz at 1484\,MHz. We thus expect the scintillation bandwidth to be detectable in B1 (raw-voltage data), while for the B2 data the frequency resolution (4\,MHz) is insufficient to resolve the expected frequency scale. If there are multiple scattering screens along the line-of-sight \citep[see, e.g.,][]{sammons_2023_mnras}, the corresponding characteristic frequency scales of the different screens should be present in the ACF (if there is sufficient resolution to resolve them). For B1, we produced filterbank data sets from the raw-voltage data recorded at Effelsberg, with different frequency resolutions, and computed the autocorrelation functions (ACFs) using an implementation of \texttt{fftconvolve} from the scipy.signal package. We then fit a Lorentzian function to ACFs of the normalized, time-averaged FRB spectrum to measure the scintillation bandwidth \citep[defined as the half-width at half-maximum of this Lorentzian fit; see, e.g.,][]{rickett_1990_araa}. We found no evidence for a frequency scale around the frequency lag corresponding to the expected scintillation bandwidth ($\sim$1.35\,MHz). Instead, we measure a scintillation bandwidth of approximately 0.1\,MHz, more than an order-of-magnitude lower than the expectation. The ACFs of B1 at different frequency resolutions are shown in Figure~\ref{fig:R12_freqacf}.

\begin{figure}
    \centering
    \includegraphics[width=0.5\textwidth]{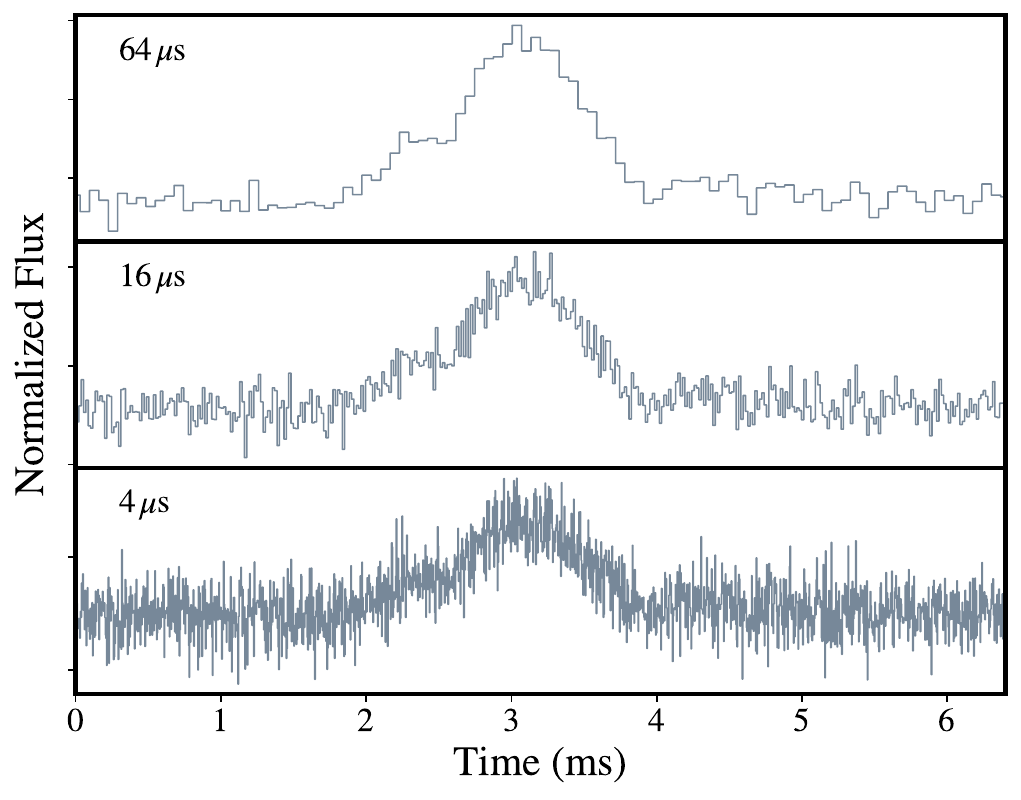}
    \caption{The frequency-averaged time profile of B1, at different time resolutions. In each case the burst has been coherently dedispersed to a DM of 580.1\,pc\,cm$^{-3}$. }
    \label{fig:E1_hitime_res}
\end{figure}

The lack of substantial scattering and scintillation in the bursts from \rtwelve suggest the bursts are not heavily modulated by propagation effects other than dispersion. Motivated by the presence of sub-millisecond time structure in bursts from other repeating FRBs \citep[e.g.,][]{nimmo_2021_natas,nimmo_2022_natas,hewitt_2023_mnras}, we investigate to see if these \rtwelve bursts contain such structure. Again we generated filterbank data sets, coherently dedispersing to a DM of 580.1\,pc\,cm$^{-3}$ (see Section~\ref{sec:dm}), with time resolutions of 64, 16 and 4\,$\upmu$s. As can be seen in Figure~\ref{fig:E1_hitime_res}, we find no significant temporal structure when probing B1 at these time resolutions. 

Interestingly, while B1 and B2 differ by about a factor of two in duration, both have similar temporal morphology, showing a small shoulder on the leading side of the burst. 

\subsection{Polarimetry: Effelsberg}
\label{sec:pol_eff}

Following \cite{nimmo_2021_natas}, we use a test pulsar (the 5-minute scan of PSR~B2255+58 in this case) instead of noise diode data to calibrate the polarimetry of our observations. Since the RM and levels of linear and circular polarization of the pulsar are confidently measured \citep[see, e.g.,][]{seiradakis_1995_aas}, we determine the calibration solutions required to reproduce the polarimetric properties of the test pulsar and then also apply these solutions to the FRB data. 

\begin{figure}
    \centering
    \includegraphics[width=0.5\textwidth]{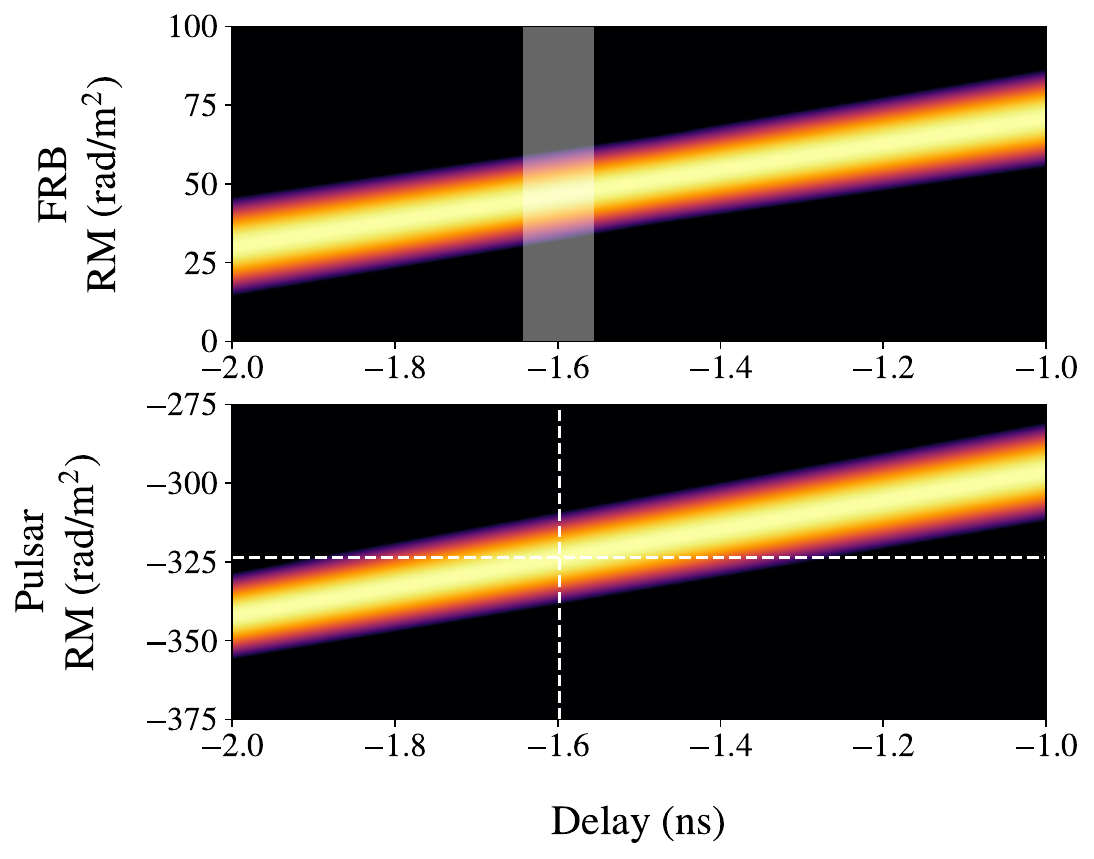}
    \caption{The RM spectra of \rtwelve (\textbf{Top}) and PSR~B2255+58 (\textbf{Bottom}) as a function of instrumental delay. The colour map has been saturated at the normalized 90\,\% power level. The horizontal dashed line in the bottom panel indicates the known RM of the pulsar, and the vertical line the corresponding delay value. In the top panel, the shaded white region shows the range of delays corresponding to a 2\,rad\,m$^{-2}$ deviation in pulsar RM. }
    \label{fig:Eff_RMs}
\end{figure}

Our uncalibrated pulsar data showed negligible levels of circular polarization, in contrast to $\sim5$\,\% of circular polarization seen from the known profile \citep[e.g.,][]{gould_1998_mnras}. We interpreted this as leakage between the two polarization hands, since the Effelsberg receiver has a circular basis (Stokes~V = RR $-$ LL). By systematically exploring a range of values, we determined that the leakage correction required to reproduce the known circular polarization profile was approximately 5\,\%. After correcting for this, we ignore second-order effects and assume the only two factors that affect Stokes~Q and U are Faraday rotation and a frequency-independent delay between the polarization hands. We performed a brute-force search for the instrumental delay, given the known RM of the pulsar \citep[$-323.5$\,rad\,m$^{-2}$; ATNF Pulsar Catalogue\footnote{\url{www.atnf.csiro.au/research/pulsar/psrcat/}};][]{manchester_2005_aj}, and found a best-fit delay of $-1.598$$\pm$0.03\,ns. To illustrate the dependency between delay and RM, we generated Faraday spectra (RM range $-2,000$ to 2,000\,rad\,m$^{-2}$) for a range of delays between $-20$ and 20\,ns. In Figure~\ref{fig:Eff_RMs} we show a region of this parameter space zooming in on the area surrounding the known RM of our pulsar and its corresponding delay.

We proceeded to calibrate the FRB data, first applying the aforementioned leakage correction, before generating the same RM-delay plot for the burst B1. Given our best-delay found from analysing the pulsar, the corresponding optimal RM for B1 is $+46.5\pm16.5$\,rad\,m$^{-2}$. To estimate the uncertainty on the RM of B1, we first assume a potential deviation of $\pm$2\,rad\,m$^{-2}$ in the RM of the pulsar (nearly double the expected ionospheric contribution). This translates to a small range of delays, indicated by the white shaded region in the top panel of Figure~\ref{fig:Eff_RMs}, which in turn translates to a range of potential RMs for B1. We took this range of RMs (measured at the 90\,\% power level of the Faraday spectra) as the final error estimate. 

After correcting the FRB data for this optimal RM, we calculated the linear and circular polarization levels. As for many repeater bursts \citep[e.g.,][ and references therein]{nimmo_2022_natas}, the burst is nearly 100\,\% linearly polarized, shows negligible levels of circular polarization and has a flat PPA across its duration (Figure~\ref{fig:portraits}).

\subsection{Polarimetry: NRT}
\label{sec:pol_nrt}
To determine the polarization properties and RM of burst B2, we calibrated both the burst and a short observation of the pulsar PSR~B0031$-$07 with \texttt{PSRCHIVE} tools \citep{hotan_2004_pasa}. We used the \texttt{pac} command with the complete \texttt{RECEPTION} calibration model \citep{vanstraten_2004_apjs,vanstraten_2013_apjs}, which also takes into account the ellipticities and orientations of the polarization hands to model the response of a receptor with non-ideal feeds. Our implementation makes use of a polarization calibration modelling (pcm) file, which is generated from a 1-h observation of the pulsar J0742$-$2822 during which the receiver-horn is rotated by 180$^{\circ}$. This mimics wide parallactic angle variation allowing for the implementation of the \texttt{RECEPTION} model \citep{guillemot_2023_aa}. 

The polarization properties of PSR~B0031$-$07 slightly vary across our NRT observing window. Nevertheless, our calibration enabled us to reproduce the polarization profile at 1369\,MHz \citep{johnston_2018_mnras} and 1642\,MHz \citep{gould_1998_mnras}. Using \texttt{PSRCHIVE}'s \texttt{rmfit} we found a RM of $9.40\pm3.75$\,rad\,m$^{-2}$ for PSR~B0031$-$07, consistent with the known value of 9.89\,rad\,m$^{-2}$ \citep[][]{manchester_2005_aj}. Having confirmed our calibration was successful, we applied the same technique to burst B2, resulting in an RM of $-5.2\pm4.9$\,rad\,m$^{-2}$.  The resulting polarization profile after correcting for this RM is shown in panels a and b of Figure~\ref{fig:portraits}b. As with B1, the burst is nearly 100\,\% linearly polarized, nearly zero percent circularly polarized and has a constant PPA.

\subsection{Constraints on Frequency-Dependent Activity}

\begin{table*}
    \centering
    \caption{Technical Specifications and Detection Rates of Different Monitoring Campaigns on \rtwelve. }
    \begin{tabular}{lccc}
    \hline
         & CHIME/FRB$^a$ & Effelsberg & NRT \\ \hline
        Total observation time$^b$ (hr) & - & 65.59 & 40.42 \\
        Fluence threshold (Jy\,ms) & 5 & 0.13 & 0.17 \\
        Frequency range (MHz) & $400-800$ & $1254-1510$ & $1228-1740$ \\
        Number of bursts detected & 15 & 1 & 1 \\
        Burst rate$^{c}$ ($\times10^{-2}$\,bursts\,hr$^{-1}$) & $4.34^{+3.81}_{-2.30}$ & $1.5_{-1.5}^{+5.7}$ & $2.5_{-2.4}^{+9.3}$ \\

        \hline
    \end{tabular} \\
    $^a$ Values from \citet{chime_2023_apj}.\\
    $^b$ For Effelsberg and the NRT, these are the total observation time of \rtwelve from PRECISE and \'ECLAT campaigns, respectively, up until the end of 2023.\\
    $^c$  These rates are not calculated over the same time period.\\ 

    \label{tab:stat_spec_ind}
\end{table*}

In Table~\ref{tab:stat_spec_ind} we tabulate the fluence thresholds, total observing time and burst rates using Poissonian statistics for the PRECISE, \'ECLAT and CHIME/FRB \citep{chime_2023_apj} observations. We note here that the repetition rates for the different telescopes are not calculated over the same time period (Figure~\ref{fig:R12_timeline}). We calculated the fluence thresholds, $F_{\text{th}}$, by assuming a $t=1$\,ms burst and using the following adaptation of the radiometer equation:

\begin{equation}
F_{\text{th}} = \text{S/N} \times \text{SEFD} \times \sqrt{\frac{t}{n_{\text{pol}} \Delta \nu }}
\label{eq:radiometer}
\end{equation}

Here, S/N is the signal-to-noise threshold used in burst searches, SEFD is the system equivalent flux density (approximately 25\,Jy for the NRT, and 14\,Jy for Effelsberg at $\sim1.4\,$GHz), $n_{\text{pol}}$ is the number of polarisations recorded (2 in these cases), and $\Delta\nu$ is the observing bandwidth.

At first glance, the burst rates might seem comparable given the large uncertainties, however, there is a more than an order-of-magnitude difference in the fluence thresholds between low- and high-frequency detections, and the bursts presented here are below the fluence threshold of CHIME/FRB. 

To better compare the burst rates at different frequencies, we calculate the statistical spectral index, $\alpha_{\rm s}$, following the formalism presented in \cite{houben_2019_aa} and also adopted in \cite{chawla_2020_apjl}. Assuming that burst energies at a given frequency ($\nu$) follow a statistical distribution that can be described by a power-law, the frequency-specific differential energy distribution of an FRB source is then given by dN$(\nu)$/dE = A$(\nu)$\,E$^\gamma$, where the power-law index, $\gamma$, is assumed to be constant over frequency and time. The statistical spectral index, $\alpha_{\rm{s}}$, is the power-law index relating the normalisation factor A$(\nu)$\ of the differential energy distribution of an FRB source at different frequencies.

Taking into account the fluence thresholds of two different instruments, $F_{\nu_1,\rm{min}}$ and $F_{\nu_2,\rm{min}}$, operating at two different frequencies, $\nu_1$ and $\nu_2$, the burst repetition rates, $\lambda_1$ and $\lambda_2$, are related as follows:

\begin{equation}
    \frac{\lambda_1}{\lambda_2}=\left(\frac{\nu_1}{\nu_2}\right)^{-\alpha\gamma}\left(\frac{F_1}{F_2}\right)^{\gamma+1}
\end{equation}

Importantly, this assumes that the repetition rates are constant with time. To then calculate $\alpha_{\rm{s}}$, we performed 10,000 trials, sampling different values for the repetition rates and $\gamma$ and solving for $\alpha_{\rm{s}}$. The repetition rates were sampled within the 90\,\% confidence intervals, and $\gamma$ from a Gaussian distribution with mean $-2.5$ and standard deviation 0.5, which contains the wide range of values of $\gamma$ measured in the literature \citep[e.g.,][]{houben_2019_aa,macquart_2019_apjl,chawla_2020_apjl,lanman_2022_apj}   . This resulted in distributions of $\alpha_{\rm{s}}$, from which we determined the 95\,\% confidence interval. For \rtwelve, we find $\alpha_{\rm{s,NRT/CHIME}}=-2.30\pm0.46$ and
$\alpha_{\rm{s,EFF/CHIME}}=-2.96\pm0.49$,
using the detections from \'ECLAT and PRECISE observations, respectively, as referenced to the CHIME/FRB detections. 

These values are largely consistent within errors to what has been estimated for \rone in the comparable $1.2-3.5$\,GHz frequency range by \citet{houben_2019_aa}, and only negligibly steeper than that for \rthree in the slightly lower $0.3-0.8$\,GHz frequency range by \citet{chawla_2020_apjl}. Notably, the sign of $\alpha_{\rm{s}}$ is consistently negative for all three of these repeating sources, implying decreasing bursting activity with increasing frequency. Future measurements using detections spanning wider frequency ranges can inform us if this index remains constant over the whole range of frequencies that the sources are detected at, or if any flattening and/or steepening is observed at relatively lower and/or higher frequencies.

\bibliography{frb}{}
\bibliographystyle{aasjournal}



\end{document}